\documentclass[conference,letterpaper]{IEEEtran}
\IEEEoverridecommandlockouts

\usepackage[letterpaper, top=1.92cm, bottom=3.2cm, left=1.92cm, right=1.92cm]{geometry}
\usepackage{pifont}
\usepackage{graphicx}
\usepackage{fancyhdr}
\usepackage{amsmath,amssymb,amstext,amsfonts}
\usepackage{siunitx}
\usepackage{bm}
\usepackage{algorithm}
\usepackage{array}
\usepackage{cite}
\usepackage{epstopdf}
\usepackage{subfigure}
\usepackage{balance}
\usepackage{algorithmicx}
\usepackage{algpseudocode}
\usepackage{epsfig}
\usepackage{braket}
\usepackage{tabulary}
\usepackage{diagbox}
\usepackage{amsthm}
\usepackage{color}
\usepackage{verbatim}
\usepackage{multirow} 
\usepackage{subfigure}
\usepackage{booktabs}  
\usepackage{siunitx}    
\usepackage{wrapfig}
\usepackage{url}
\usepackage[normalem]{ulem}
\usepackage{comment}
\def\BibTeX{{\rm B\kern-.05em{\sc i\kern-.025em b}\kern-.08em
    T\kern-.1667em\lower.7ex\hbox{E}\kern-.125emX}}
\begin{document}

\title{CIG-MAE: Cross-Modal Information-Guided Masked Autoencoder for Self-Supervised WiFi Sensing}

\author{Gang~Liu, Yanling~Hao, and~Yixuan~Zou%
\thanks{(Corresponding author: Yanling Hao.)}
\thanks{Gang Liu and Yixuan Zou are with the School of Electronic Engineering and Computer Science, Queen Mary University of London, London E1 4NS, U.K. (e-mail: gangliu2004@outlook.com; yixuan.zou@qmul.ac.uk).}%
\thanks{Yanling Hao is with the School of Computing and Engineering, University of West London, London W5 5RF, U.K. (e-mail: yanling.hao@uwl.ac.uk).}%
\thanks{Copyright (c) 2026 IEEE. Personal use of this material is permitted. However, permission to use this material for any other purposes must be obtained from the IEEE by sending a request to pubs-permissions@ieee.org.}%
}

\maketitle

\begin{abstract}
Human Action Recognition using WiFi Channel State Information (CSI) has emerged as an attractive alternative to vision-based methods due to its ubiquity, device-agnostic nature, and inherent privacy-preserving capabilities. However, the high cost of manual annotation and the limited scale of publicly available CSI datasets restrict the performance of supervised approaches. Self-supervised learning (SSL) offers a promising avenue, but existing contrastive paradigms rely on data augmentations that conflict with the physical semantics of radio signals and require large-batch training, making them poorly suited for CSI. To overcome these challenges, we introduce CIG-MAE—a Cross-Modal Information-Guided Masked Autoencoder—that reconstructs both the amplitude and phase of CSI using a symmetric dual-stream architecture with a high masking ratio. Specifically, we propose an Adaptive Information-Guided Masking strategy that dynamically allocates attention to time–frequency regions with high information density to improve learning efficiency, and we incorporate a Barlow Twins regularizer to align cross-modal representations without negative samples. Experiments on three public datasets show that CIG-MAE consistently outperforms SOTA SSL methods and even surpasses a fully supervised baseline, demonstrating superior data efficiency, robustness, and representation generalization.
\end{abstract}

\section{Introduction}\label{sec:intro}
Human Action Recognition (HAR) has garnered significant attention for its broad applications in health monitoring, elderly care, and smart homes \cite{Intro_DailyAcMonitoring_2023,Intro_SignalDetecting_2019,Intro_HealthMonitoring_2020,Intro_FallDetection_2017,Intro_ElderlyCare_2021}. Among various sensing modalities, HAR based on WiFi Channel State Information (CSI-HAR) has emerged as a particularly promising approach due to its ubiquity, device-agnostic nature, privacy-preserving characteristics, and capability to operate in non-line-of-sight (NLOS) environments \cite{Intro_Ubiquitous_2022,Intro_wifi_2023}.

Despite its potential, the practical deployment of CSI-HAR is significantly hampered by the high cost and difficulty of acquiring large-scale, accurately labeled datasets. The performance of conventional supervised learning models is fundamentally constrained by data scarcity and the challenge of covering diverse subjects and environments, which can lead to poor generalization and domain shift issues \cite{Intro_LabelDifficult_2024,Intro_DomainShift_2018}. To mitigate this dependency on labeled data, Self-Supervised Learning (SSL) has been introduced as a powerful paradigm capable of leveraging large volumes of unlabeled CSI to learn transferable representations and enhance robustness in unseen scenarios \cite{Intro_SSLSurvey_2024,Intro_SSLSurveyWifi_2023}.
\begin{figure}[H]
\centering
\subfigure[Amp Reconstruction Heatmap\label{AmpHeat}]
{\includegraphics[width=0.48\columnwidth]{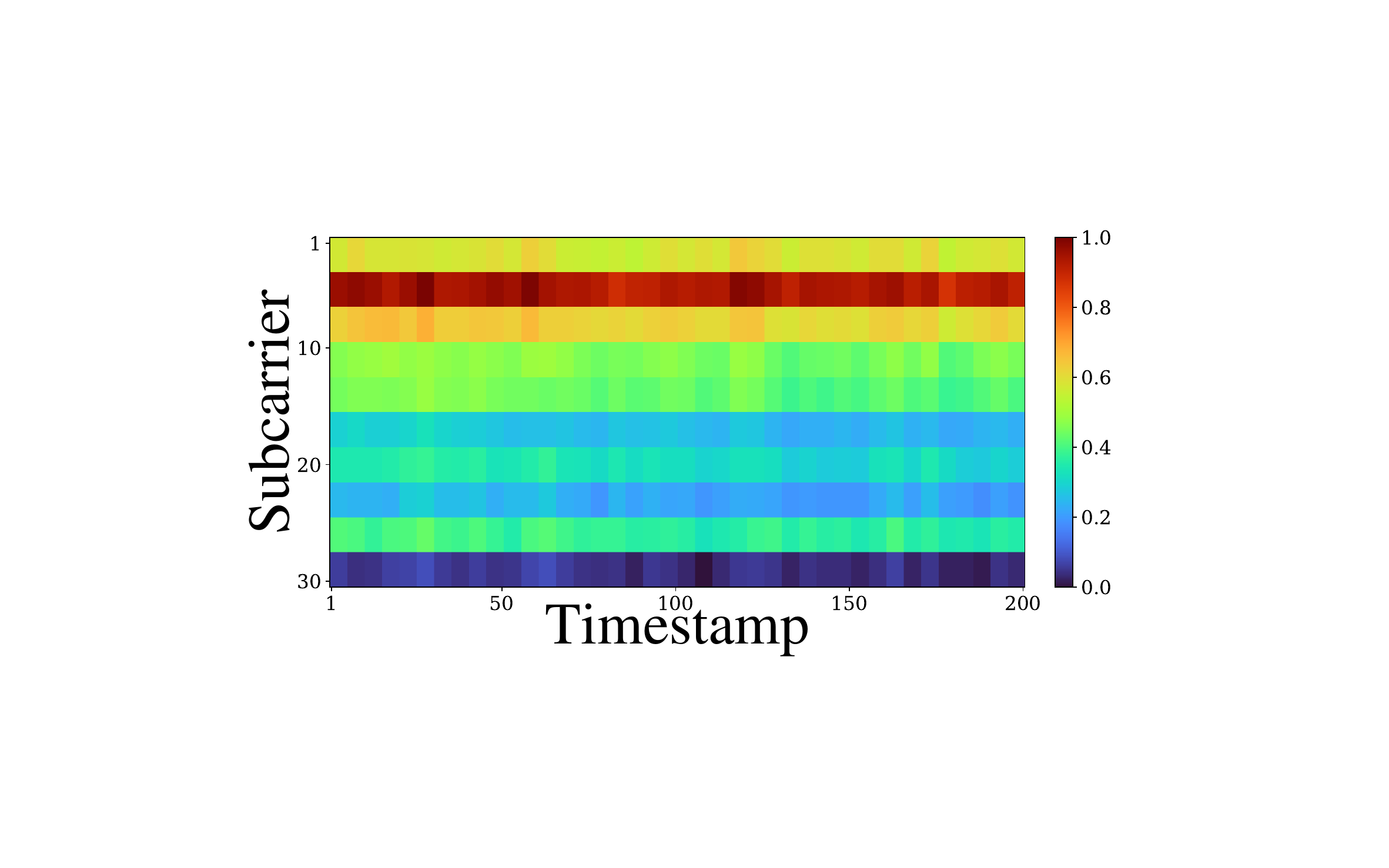}}\hspace{2mm}
\subfigure[Phase Reconstruction Heatmap\label{PhaseHeat}]
{\includegraphics[width=0.48\columnwidth]{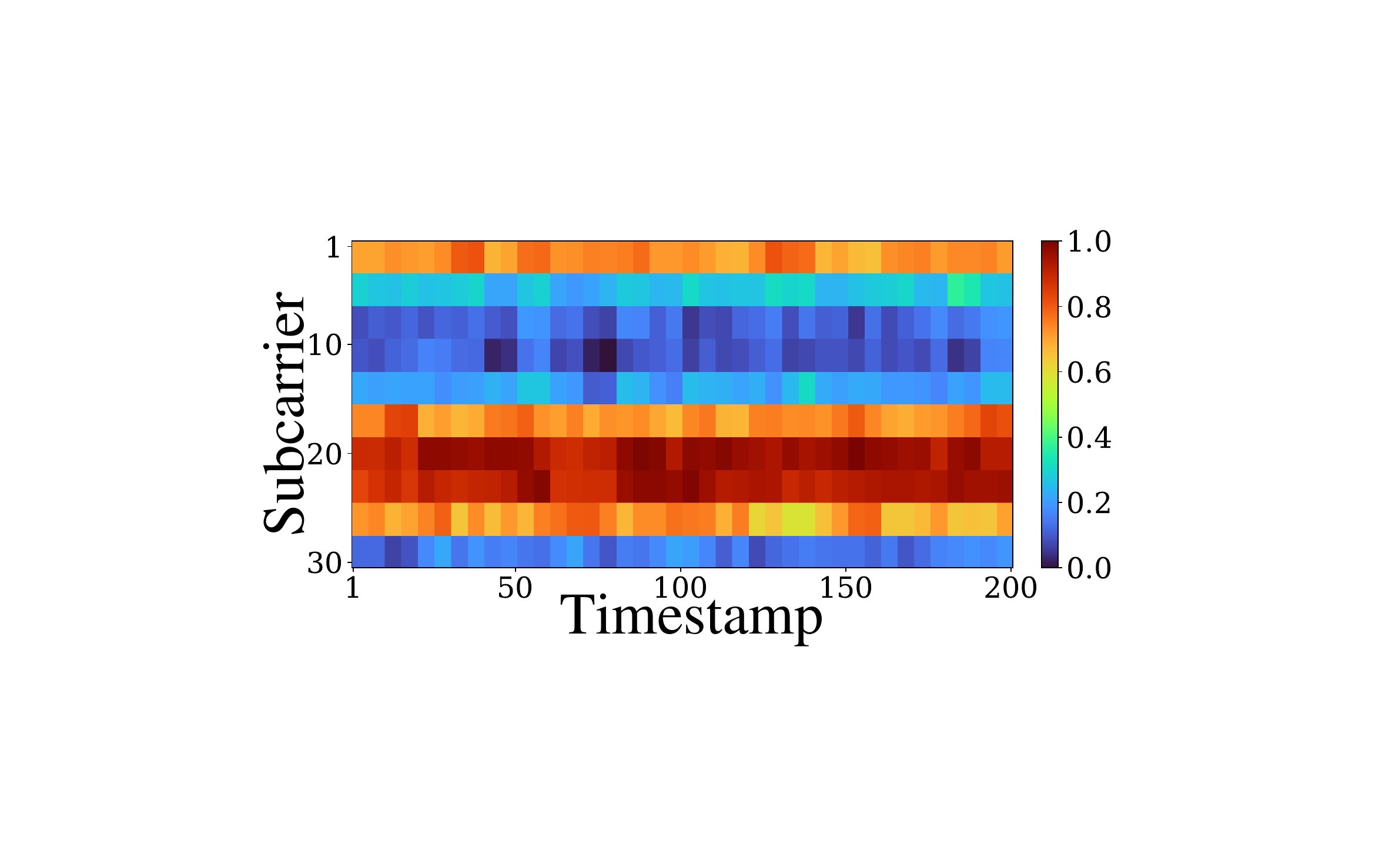}}
\caption{Reconstruction error heatmaps for amplitude and phase on the subcarrier-time plane. A larger reconstruction error (warmer color) is interpreted as higher information density. Both modalities show high difficulty on specific subcarrier indices, but their high-difficulty regions do not fully overlap, reflecting modal complementarity and non-uniform information distribution.}
\label{heat_map}
\end{figure}
However, mainstream SSL paradigms, predominantly developed for computer vision and natural language processing, often prove suboptimal when directly applied to WiFi CSI data. In particular, contrastive learning depends heavily on data augmentations that lack consistency with radio propagation physics and typically requires large-batch training, which is incompatible with the small-dataset reality of WiFi sensing. These mismatches can corrupt semantic structure and undermine the ability to learn invariant features.

A key factor that has been largely overlooked is the intrinsic complementary nature of the amplitude and phase components of CSI. While much existing SSL work relies solely on amplitude \cite{SSL_DiffusionCLAR_Wifi_2024,SSL_MIM_2024}, phase has been shown to be highly sensitive to micro-motion and propagation geometry. Yet, even when phase is used, it is often treated as a unidirectional prediction target rather than a jointly learned modality, failing to capture their inherent coupling \cite{SSL_Autosen_Wifi_2024}. Moreover, as illustrated in Fig.~\ref{heat_map}, the information density—interpreted as reconstruction difficulty—exhibits strong non-uniformity across the time–frequency axis for both modalities, and critically, their high-density regions only partially overlap. This indicates that amplitude and phase provide complementary physical information and that a well-designed SSL framework must explicitly model and leverage this complementarity to obtain discriminative and physically meaningful representations.

To address these challenges, we propose the Cross-Modal Information-Guided Masked Autoencoder (CIG-MAE), a self-supervised framework tailored to the physical structure and sensing properties of CSI. CIG-MAE adopts a symmetric dual-stream architecture that jointly reconstructs amplitude and phase under a high masking ratio, eliminating the need for problematic augmentations and large batch sizes. Furthermore, we introduce an Adaptive Information-Guided Masking (AIM) mechanism that allocates visibility to time–frequency regions with higher information density, thereby enhancing representation learning efficiency and boosting model performance. The main contributions of this work are summarized as follows:
\begin{itemize}
    \item We propose a symmetric dual-stream self-supervised reconstruction framework, CIG-MAE, tailored to the physical properties of CSI. This framework reconstructs the amplitude and phase components of CSI with a high masking ratio and introduces a non-contrastive Barlow Twins (BT) regularizer to align the dual-stream representations and eliminate feature redundancy. This design explicitly models the complementarity between amplitude and phase, effectively avoiding the reliance of existing methods on data augmentations that are inconsistent with physical semantics and on large-batch configurations.
    \item We design an AIM strategy to enhance data efficiency and representation quality. This strategy introduces a learnable policy network that can evaluate the information value of each time-frequency region, thereby dynamically focusing the model's "visibility budget" on regions with higher information density that are more critical for learning discriminative features, addressing the inefficiency of random masking from a mechanistic perspective.
    \item We conduct systematic evaluations on three public datasets. The experimental results show that under the standard linear probing protocol, the performance of the proposed CIG-MAE not only significantly surpasses various advanced self-supervised baselines but even exceeds a fully supervised model in certain scenarios. Furthermore, exhaustive ablation studies and parameter analyses have verified the effectiveness of our model's components and its overall robustness.
\end{itemize}

The remainder of this paper is organized as follows: Section~\ref{sec:related} reviews related work. Section~\ref{sec:method} elaborates on our proposed CIG-MAE framework. Section~\ref{sec:experiments} reports detailed experimental settings and results. Section~\ref{sec:discussion} discusses broader implications and potential extensions of CIG-MAE. Finally, Section~\ref{sec:conclusion} concludes the paper and discusses future research directions.

\section{Related Work}\label{sec:related}
SSL has achieved remarkable success across various domains \cite{SSL_BERT_2019,SSL_MAE_2021,SSL_BEIT_2022,SSL_Speech_2022} and is gradually being adopted for signal processing tasks \cite{SSL_maskCAE_2024,SSL_SenseAndLearning_2021,SSL_RFURL_2022}. Early research on SSL for HAR explored various pretext tasks. Saeed et al. \cite{SSL_SenseAndLearning_2021} systematically designed eight different pretext tasks such as feature prediction and transformation recognition for diverse signals. Subsequently, they extended SSL to federated settings \cite{SSL_FederatedscalogramLearning_2021}, proposing scalogram-signal correspondence learning via wavelet transforms. As research deepened, methods specifically tailored to the physical structure of WiFi CSI have emerged, broadly categorized into discriminative (contrastive/predictive) and generative paradigms.
\subsection{Contrastive and Predictive Paradigms}
The fundamental idea of contrastive learning is to maximize representation consistency between different views of the same sample while minimizing consistency between different samples \cite{SSL_MoCo_2020,SSL_BT_2021,SSL_VICREG_2022,SSL_SimCLR_2020}. In WiFi sensing, current works optimize this paradigm primarily through view construction and objective function design.

Regarding \textit{view construction}, Lau et al. \cite{SSL_2ReceiversContrastive_2022} treated CSI from spatially separated receivers as positive pairs. Liu et al. \cite{SSL_STFNet_AccWifi_2021} introduced STFNets to mine time-frequency features, demonstrating the importance of frequency domain augmentations. Song et al. \cite{SSL_RFURL_2022} focused on multi-representation consistency, utilizing a translator-predictor structure to map distinct radio frequency (RF) features like Angle of Arrival (AoA) and Doppler Frequency Shift (DFS) into a unified space. Xu et al. \cite{SSL_DualStream_Wifi_2023} employed a dual-stream architecture to collaboratively model spatial and channel features. Lyons et al. \cite{WiFiAct_SSL_2024} proposed WiFiAct, which combines an environment-invariant preprocessing pipeline with a Bayesian CNN for generalized learning, though its partial reliance on labels distinguishes it from pure SSL.

Regarding \textit{objective functions}, Yang et al. \cite{SSL_AutoFi_Wifi_2023} proposed AutoFi, which enhances consistency by minimizing the Kullback-Leibler (KL) divergence of probability distributions alongside mutual information and geometric structure constraints. Chen et al. \cite{SSL_MIM_2024} addressed the instability of KL divergence by maximizing Jensen-Shannon (JS) divergence and introducing Gaussian regularization to improve generalization. Recently, Xiao et al. \cite{SSL_DiffusionCLAR_Wifi_2024} incorporated diffusion models, specifically Denoising Diffusion Probabilistic Models (DDPM) to generate high-quality, temporally specific augmentations and designed adaptive weights based on activity information.

In predictive learning, the model learns representations that capture the dynamics and latent structures of the data by predicting future segments, missing context, or certain transformation properties \cite{SSL_TSCPC_2019,SSL_TS_Prediction_survey_2024}.  Haresamudram et al. \cite{SSL_SensorsCPC_2020} applied Contrastive Predictive Coding (CPC) to wearable sensors. In CSI-HAR, Barahimi et al. \cite{SSL_CAPC_2024} proposed CAPC, combining CPC with BT to leverage uplink/downlink views for capturing temporal context and view consistency.

\begin{figure}[t]
\centering
\subfigure[]{\includegraphics[width=0.18\columnwidth]{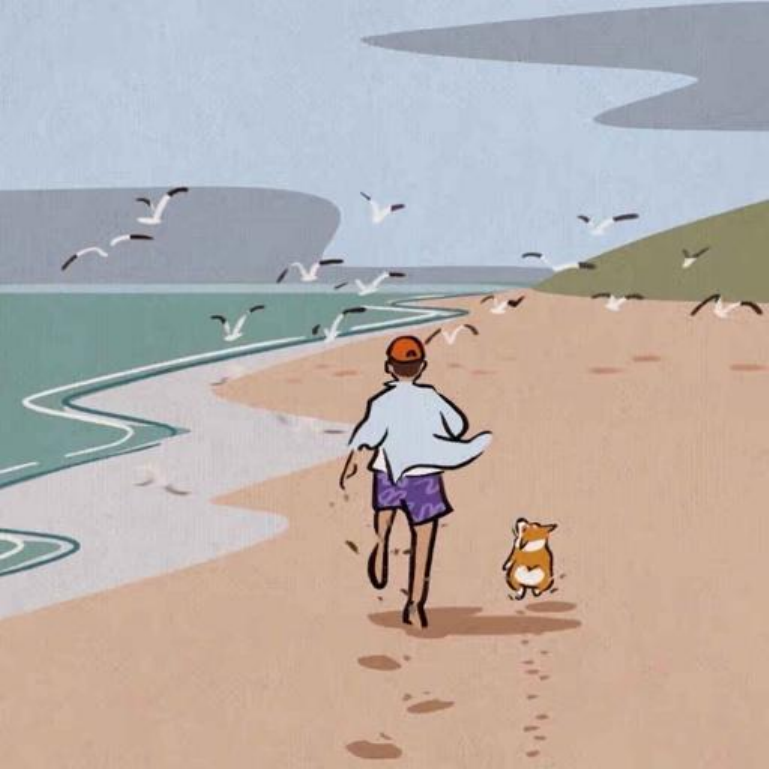}}\hspace{1mm}
\subfigure[]{\includegraphics[width=0.18\columnwidth]{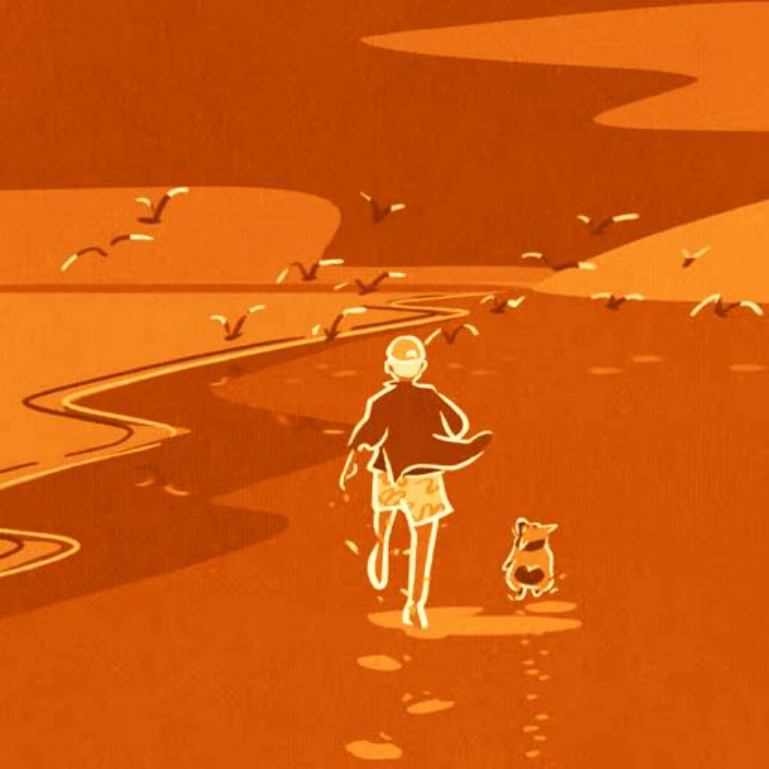}}\hspace{1mm}
\subfigure[]{\includegraphics[width=0.18\columnwidth]{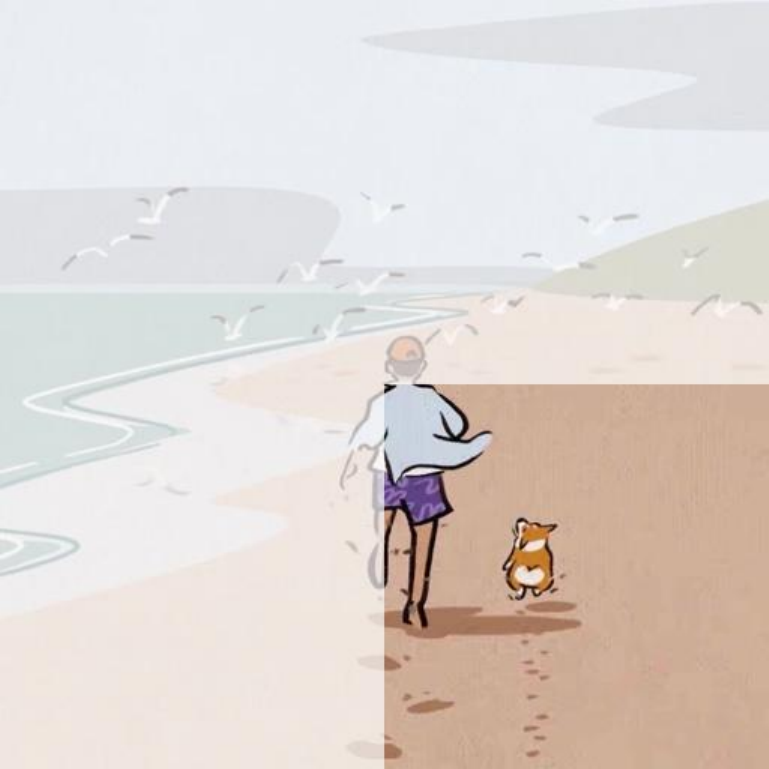}}\hspace{1mm}
\subfigure[]{\includegraphics[width=0.18\columnwidth]{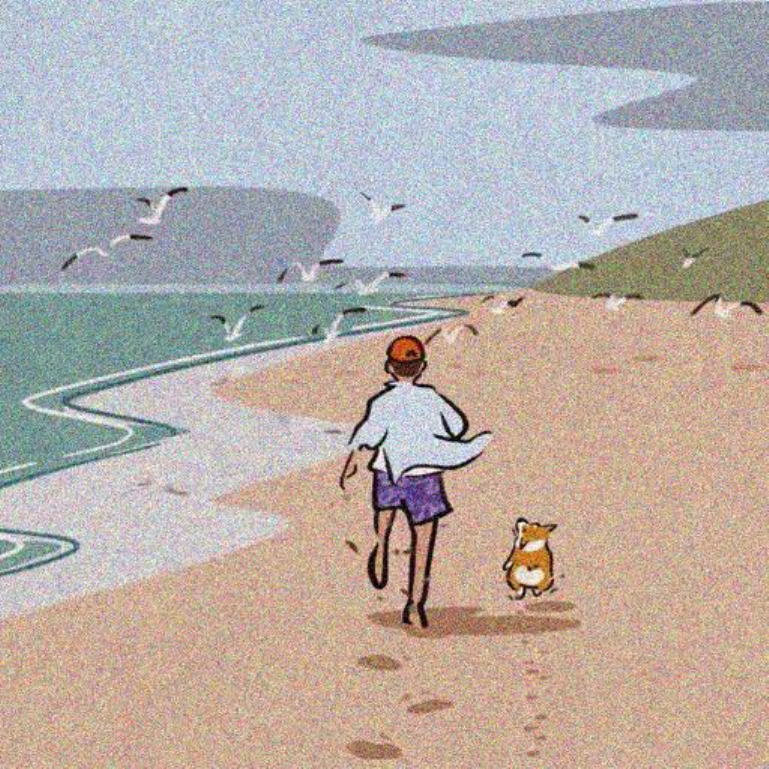}}\hspace{1mm}
\subfigure[]{\includegraphics[width=0.18\columnwidth]{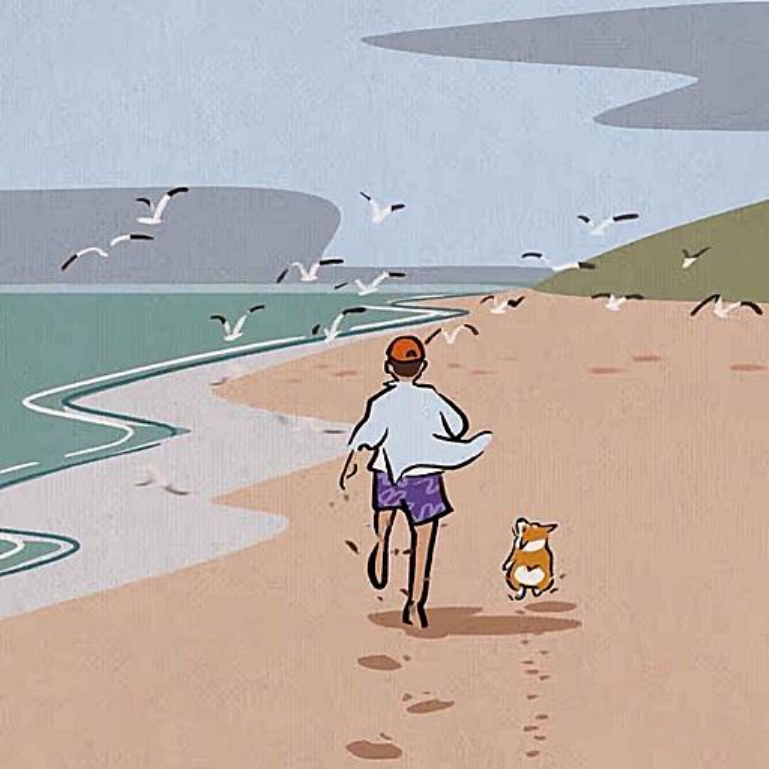}}
\vspace{0mm} 
\subfigure[]{\includegraphics[width=0.18\columnwidth]{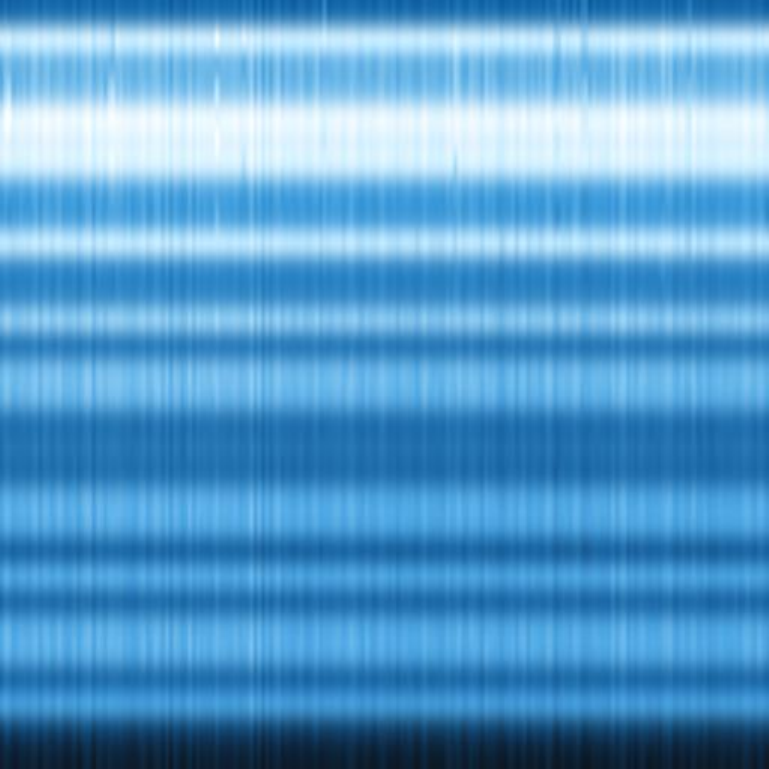}}\hspace{1mm}
\subfigure[]{\includegraphics[width=0.18\columnwidth]{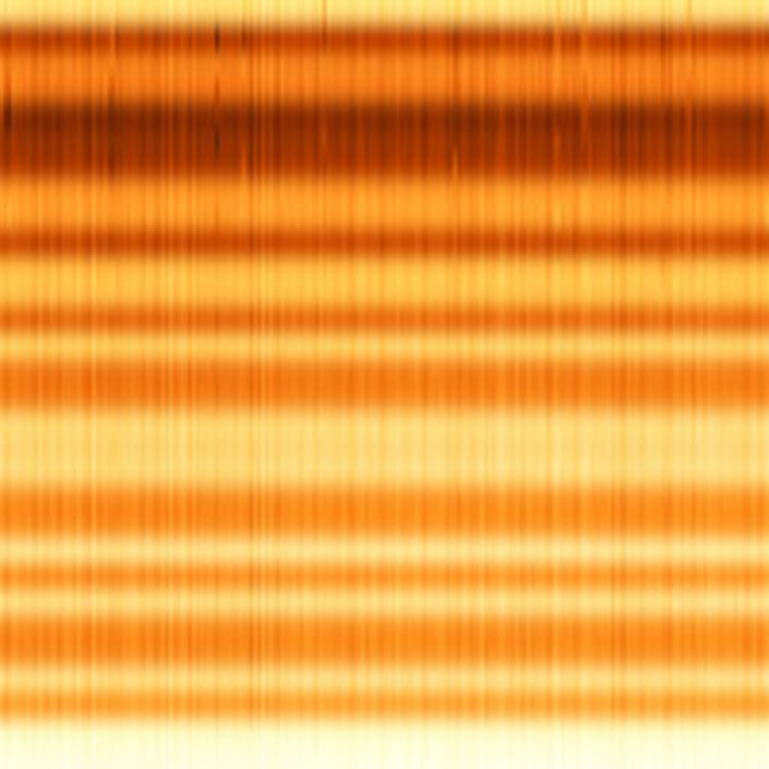}}\hspace{1mm}
\subfigure[]{\includegraphics[width=0.18\columnwidth]{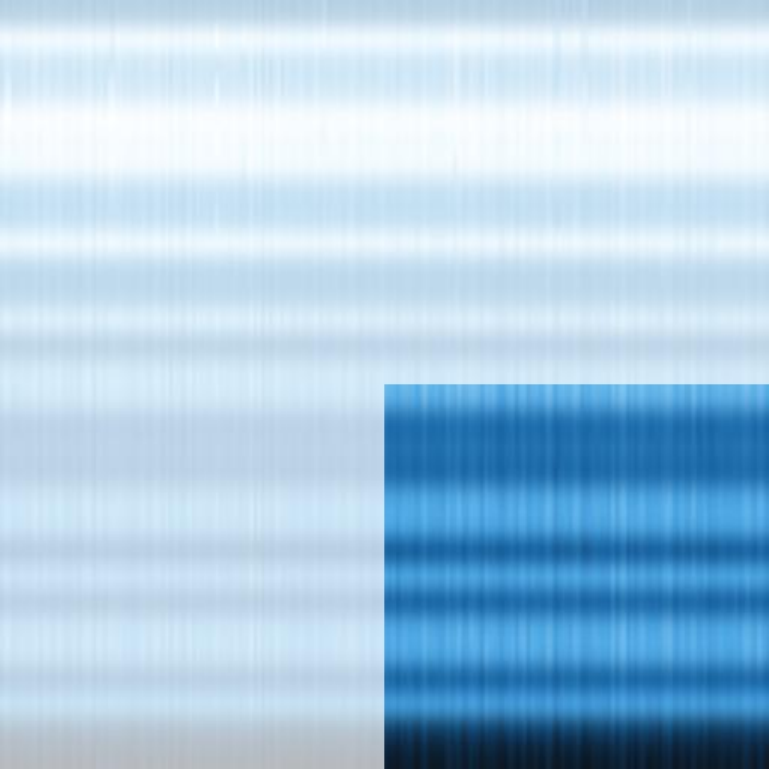}}\hspace{1mm}
\subfigure[]{\includegraphics[width=0.18\columnwidth]{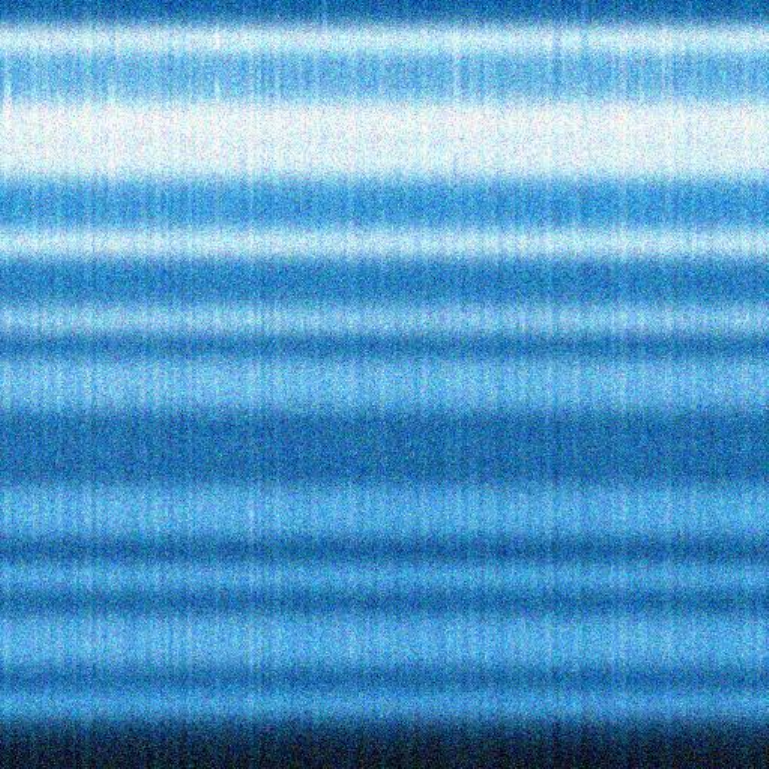}}\hspace{1mm}
\subfigure[]{\includegraphics[width=0.18\columnwidth]{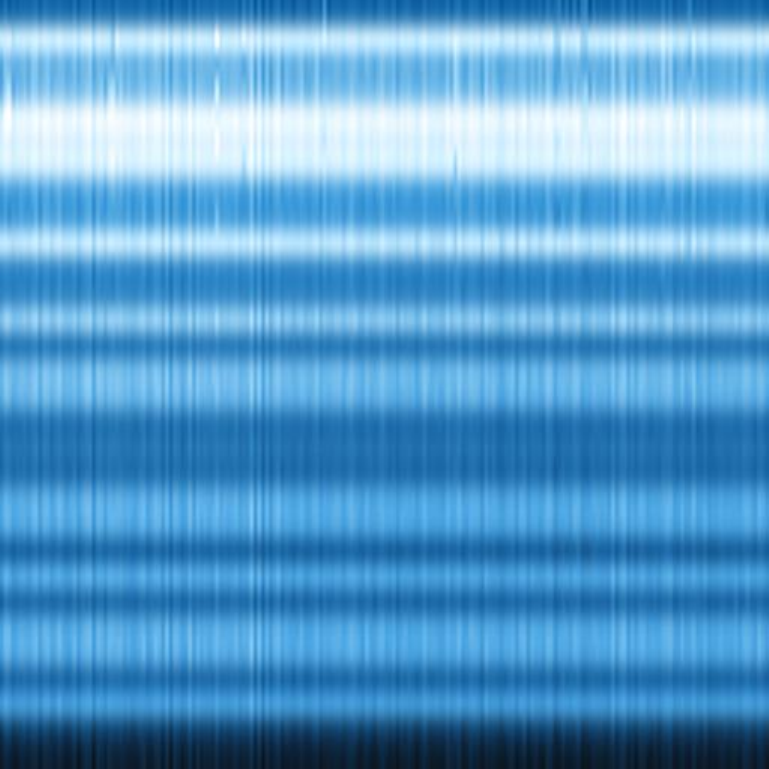}}
\caption{Illustration of five visualization/augmentation operators applied to a natural image (top row) and to CSI data (bottom row). From left to right: Original, Color Mapping, Crop, Gaussian Noise, and Unsharp Mask. Directly transferring vision augmentations to CSI can misalign with RF physical semantics.}
\label{augmentation_comparison}
\end{figure}

However, these discriminative paradigms face two inherent limitations in the CSI context. First, they rely heavily on view construction via augmentations (e.g., cropping, noise injection) that are often semantically inconsistent with the physical properties of radio signals \cite{CSIAugBasedonImageMethod,imageAugIsDifferentWithCSI_1}, as illustrated in Fig.~\ref{augmentation_comparison}. While solutions like using multiple receivers or specialized augmentations exist \cite{SSL_RFURL_2022,SSL_2ReceiversContrastive_2022,SSL_CAPC_2024,imageAugIsDifferentWithCSI_2}, they often introduce extra hardware costs or complexity. Second, contrastive learning typically requires large batches and numerous negative samples to perform well \cite{SSL_SimCLR_2020,SSL_MoCo_2020}, a condition that large-scale image datasets like ImageNet can satisfy \cite{ImageNet}. This is fundamentally at odds with the CSI domain, where public datasets are typically small, containing only hundreds to thousands of samples \cite{SSL_Survey_CST_2025}.  Predictive methods, which learn by predicting future or masked segments \cite{SSL_CAPC_2024}, often still rely on contrastive objectives and are sensitive to hyperparameters, magnifying training instability under small-sample, high-noise conditions.
\subsection{Generative and Reconstructive Paradigms}
Generative learning, which forces models to learn intrinsic structures by reconstructing masked inputs, offers a solution more suitable for small-sample scenarios with specific physical structures \cite{MIMdoNotNeed_Aug_and_Sample}. In general sensor domains, Cheng et al. proposed MaskCAE \cite{SSL_maskCAE_2024}, utilizing sparse convolutions to process masked sensor data without explicit augmentation. Miao et al. \cite{SSL_STMAE_2024} proposed STMAE for wearable devices, employing spatial-temporal masking and an asymmetric autoencoder to model device correlations. In WiFi-specific contexts, Yang et al. \cite{SSL_MaskFi_2024} proposed MaskFi, which transforms multi-modal (WiFi-Vision) data into discrete tokens via a VQ-VAE and reconstructs them using a shared Transformer. Ji et al. \cite{SSL_SiFall_Wifi_2022} proposed SiFall, framing fall detection as anomaly detection by using a VAE to model non-fall activities and utilizing reconstruction error as the anomaly signal. Focusing on modal correlation, Gao et al. \cite{SSL_Autosen_Wifi_2024} proposed AutoSen, a cross-modal autoencoder that reconstructs phase from amplitude to capture intrinsic CSI semantics.

Despite avoiding complex augmentations, existing generative methods exhibit structural shortcomings. Approaches like MaskFi rely on heavy backbones (e.g., Transformers) that are computationally prohibitive and unstable on small, noisy CSI datasets \cite{ViTcannotworkonSmallDataset}. Asymmetric designs, such as AutoSen's unidirectional prediction, create an information bottleneck by assuming amplitude fully specifies phase. Moreover, standard uniform random masking ignores the non-uniform information density of CSI, potentially wasting learning budgets on redundant regions \cite{randomMaskIsNotEnough}. 
\section{Methodology}\label{sec:method}
\begin{figure*}[t]
    \centering
    \includegraphics[width=1\linewidth]{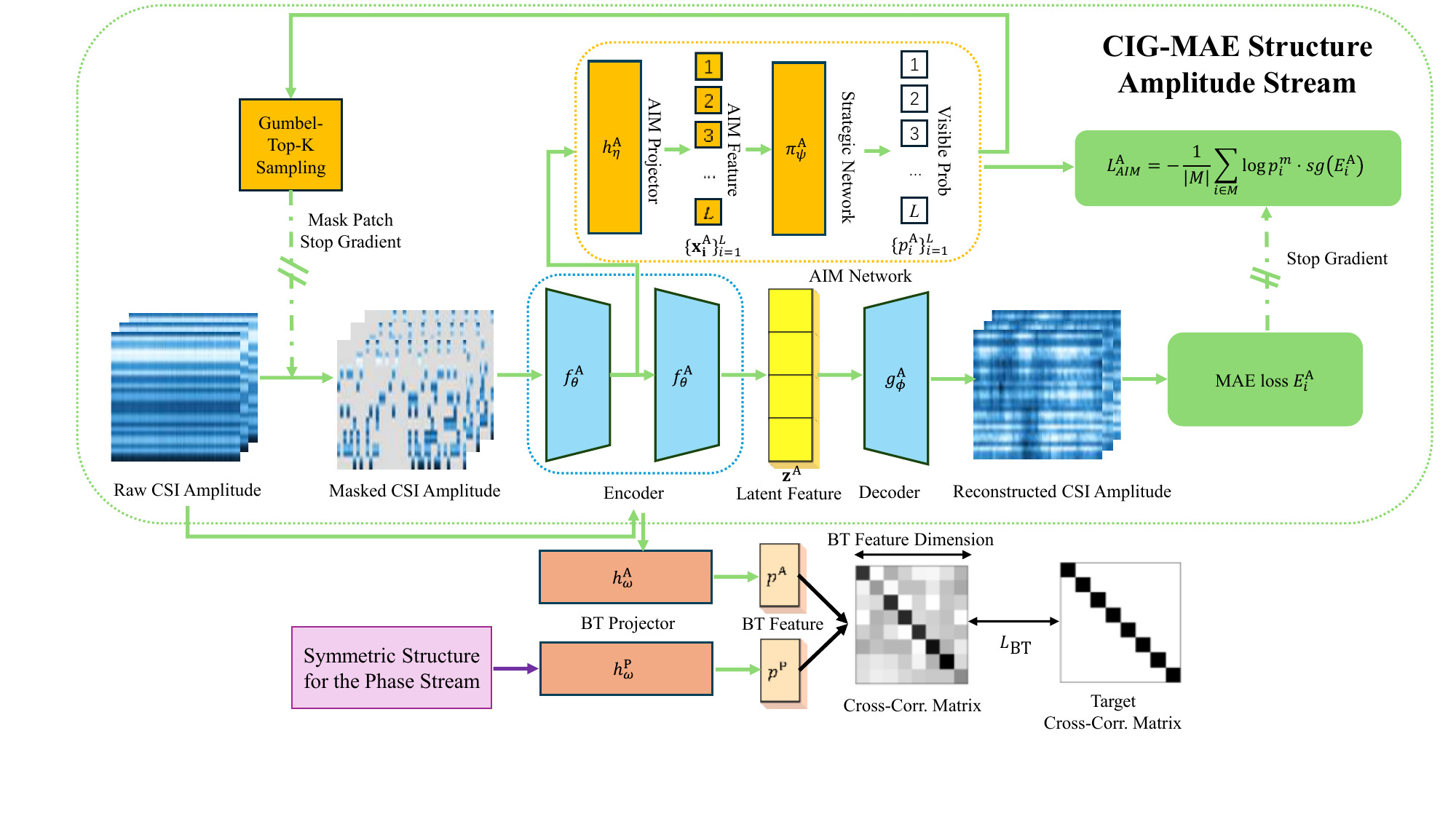}
    \caption{Schematic overview of the CIG-MAE framework. The architecture integrates three core modules corresponding to Section \ref{sec:method}:
    (1) \textbf{Adaptive Information-Guided Masking (Sec. \ref{sec:aim})}: The policy network evaluates information density to generate non-uniform masks on the $S\times T$ plane;
    (2) \textbf{Dual-Stream Convolutional Backbone (Sec. \ref{sec:reconstruction})}: Symmetric CNN encoder-decoder networks reconstruct the signal from masked amplitude and phase streams;
    (3) \textbf{Cross-Modal Representation Alignment (Sec. \ref{sec:alignment})}: A Barlow Twins regularizer aligns unmasked representations to ensure semantic consistency.}
    \label{CIG_MAE_pipeline}
\end{figure*}

To address the challenges of data scarcity and non-uniform information density in WiFi sensing, we propose CIG-MAE, a self-supervised framework tailored to the physical representation of CSI. As illustrated in Fig.~\ref{CIG_MAE_pipeline}, the proposed framework processes synchronized dual-stream CSI through a \textit{Filter-then-Reconstruct} paradigm. The pipeline is composed of three integral modules: an \textit{AIM} mechanism that actively identifies information-dense signal patches; a \textit{Dual-Stream Convolutional Backbone} that reconstructs the signal from these focused regions; and a \textit{Cross-Modal Representation Alignment} module that enforces semantic consistency between modalities. The specific design motivations and details are described below.

\subsection{Preliminaries: Signal Formulation and Properties}
\label{sec:preliminaries}
WiFi CSI characterizes the wireless channel at a fine-grained level using Orthogonal Frequency Division Multiplexing (OFDM). Mathematically, the CSI for subcarrier $k$ at time $t$ is represented as a complex value:
\begin{equation}
    H_k(t) = A_k(t) e^{j\phi_k(t)}
    \label{eq:csi_model}
\end{equation}
where $A_k(t)$ is the amplitude and $\phi_k(t)$ is the phase. Consequently, the input to our model is defined as a synchronized dual-stream tensor $(\mathbf{X}^{A}, \mathbf{X}^{P})$, where each $\mathbf{X}^{m}\in\mathbb{R}^{N\times S\times T}$ for modality $m\in\{A,P\}$, with $N$, $S$, and $T$ denoting antennas, subcarriers, and time steps, respectively.

This formulation underscores two critical physical properties that inform our architectural design. First, the coupling of $A_k(t)$ and $\phi_k(t)$ implies that single-stream models inherently discard complementary modal information, necessitating a \textbf{symmetric dual-stream architecture}. Second, human activities typically manifest as locally structured patterns on the time-frequency plane ($S \times T$). Unlike vision tasks, CSI is heavily contaminated by background. This renders standard random masking inefficient, motivating the introduction of an adaptive masking strategy.
\subsection{Adaptive Information-Guided Masking (AIM)}\label{sec:aim}
Standard masked autoencoders typically employ uniform random masking. However, given the high noise floor in CSI data, such a strategy usually fails to distinguish between information-dense and redundant regions, leading to a significant waste of computational resources on meaningless reconstruction tasks \cite{informationSparse_1,informationSparse_2}. To mitigate this, we incorporate the AIM strategy \cite{AdaMAE_CVPR_2023} to dynamically allocate visibility to regions with high information density.

\textbf{Implicit Patch Embedding.} We first discretize the input $\mathbf{X}^{m}$ into a grid of patches. We define the patch indices as $\mathcal{L}=\{1,\dots,L\}$, where each patch has dimensions $(s_p, t_p)$. To avoid the computational overhead of a separate tokenizer, we strictly align the kernel size and stride of the encoder's first convolutional layer with $(s_p, t_p)$. This yields native patch embeddings, which are projected to a policy feature space $\{\mathbf{x}^{m}_{i}\}_{i=1}^{L}$ via a linear layer $h_{\eta}^{m}$.

\textbf{Policy Network and Probabilistic Sampling.} A lightweight Policy Network $\pi_{\psi}^{m}$, implemented as a single-layer Transformer Encoder, evaluates the contextual importance of each patch. It aggregates global information via multi-head self-attention (MHA) and produces hidden states $\mathbf{U}$ via an MLP:
\begin{equation}
\mathbf{Z} = \{\mathbf{x}^{m}_{i}\} + \mathrm{MHA}(\mathrm{LN}(\{\mathbf{x}^{m}_{i}\})), \quad \mathbf{U} = \mathbf{Z}+\mathrm{MLP}(\mathrm{LN}(\mathbf{Z})).
\end{equation}
The importance probability $p^{m}_{i}$ (i.e., the probability of being \textbf{visible}) for the $i$-th patch is then derived via a softmax projection:
\begin{equation}
p^{m}_{i} = \frac{\exp(\mathbf{w}^{\top}\mathbf{U}_{i})}{\sum_{j=1}^{L}\exp(\mathbf{w}^{\top}\mathbf{U}_{j})}.
\end{equation}
Based on $\{p^{m}_{i}\}$, we employ Gumbel-Top-K sampling to partition the patch set $\mathcal{L}$ into a masked set $\mathcal{M}$ (where $|\mathcal{M}| = \rho L$, mask ratio $\rho\in(0,1)$) and a visible set $\mathcal{V}$. This generates the binary mask $\Pi$ applied to the input: $\tilde{\mathbf{X}}^{m} = \mathbf{X}^{m} \odot \Pi$.
\textbf{Policy Optimization via Reconstruction Reward.} 
The core innovation of AIM lies in how we define \textit{valuable} patches without ground-truth labels. We propose a hypothesis rooted in information theory: \textit{patches that are difficult to reconstruct from their context contain higher information density (e.g., human activity), whereas patches that are easily inferable correspond to redundant background.}

Guided by this intuition, we model the masking process as learning a mask distribution induced by the policy network. Let $q_{\psi}(\mathcal{M}\mid \mathbf{X}^{m})$ denote the distribution over the masked set $\mathcal{M}$ generated by the Gumbel-Top-K sampler from the policy scores. For a sampled mask, we define the patch-wise reconstruction error $E_i^m$ as the learning signal. Unlike adversarial approaches that simply mask salient regions, our goal is to identify high-error regions and increase their likelihood of being preserved in future iterations. Given the high masking ratio (e.g., 95\%) in WiFi sensing, random masking risks removing the sparse activity signal entirely. By explicitly encouraging the network to select difficult-to-reconstruct regions into the visible set $\mathcal{V}$, we ensure the model focuses its limited visibility budget on capturing the discriminative ``skeleton'' of the signal.

Accordingly, we optimize the policy using a REINFORCE-style objective\cite{REINFORCE_refer_1,REINFORCE_refer_2}. Under a fixed masking budget, Eq.~(\ref{eq:aim_loss}) acts as a tractable single-sample surrogate that assigns larger gradients to high-error patches:
\begin{equation}
\mathcal{L}_{\mathrm{AIM}}^{m}
=
-\frac{1}{|\mathcal{M}|}\sum_{i\in\mathcal{M}}
\log p_i^m \cdot \operatorname{stopgrad}(E_i^m),
\label{eq:aim_loss}
\end{equation}
where $p_i^m$ is the policy score used by the sampler. The $\operatorname{stopgrad}(\cdot)$ operator prevents the reconstruction backbone from artificially inflating the reward signal, so that the encoder-decoder is trained only to minimize reconstruction loss, while the policy network independently learns which regions should remain visible.

\noindent\textit{Clarification.} AIM is \emph{RL-inspired} rather than a full sequential reinforcement learning formulation. We use a REINFORCE-style score-function estimator only to optimize the discrete mask selection induced by Gumbel-Top-K sampling. AIM does not involve multi-step trajectories, delayed rewards, or an independently evolving environment. Therefore, Eq.~(\ref{eq:aim_loss}) is used directly as a synchronous reconstruction-derived training signal within the same self-supervised optimization loop, and we do not employ additional RL-specific stabilization techniques such as experience replay, baseline subtraction, or reward normalization.

\textbf{Theoretical Justification.}
We further clarify the connection between Eq.~(\ref{eq:aim_loss}) and mutual information maximization. Let $X$ denote the full input, and let $X_V$ and $X_M$ denote the visible and masked subsets under a sampled mask. Since $H(X)$ is constant for a given sample, maximizing the mutual information between the visible subset and the full signal is equivalent to minimizing the conditional uncertainty of the full signal given the visible subset:
\begin{equation}
\max I(X_V;X)
=
H(X)-H(X\mid X_V)
\propto
\min H(X\mid X_V).
\label{eq:mi_objective}
\end{equation}

Under the common patch-wise approximation, we write$H(X\mid X_V)\approx\sum_{i\in\mathcal{M}} H(X_i\mid X_V),$
where $X_i$ denotes the $i$-th masked patch. Moreover, minimizing the $L_1$ reconstruction loss is equivalent to maximum likelihood estimation under a Laplace conditional model, $X_i \mid X_V \sim \mathrm{Laplace}(\mu_i,b_i)$, for which the conditional entropy satisfies$H(X_i\mid X_V)=1+\ln(2b_i)$. Since the optimal expected $L_1$ error is proportional to $b_i$, a larger reconstruction error implies a larger conditional entropy. Therefore, maximizing $I(X_V;X)$ can be relaxed into minimizing the \emph{expected} entropy (or equivalently, reconstruction error) of the masked subset:
\begin{equation}
\max I(X_V;X)
\Longrightarrow
\min_{\psi}\;
\mathbb{E}_{\mathcal{M}\sim q_{\psi}(\mathcal{M}\mid X)}
\left[
\sum_{i\in\mathcal{M}} E_i
\right].
\label{eq:expected_mask_error}
\end{equation}
Applying the score-function estimator to Eq.~(\ref{eq:expected_mask_error}) yields
\begin{equation}
\nabla_{\psi}\mathcal{J}(\psi)
\propto
\mathbb{E}_{\mathcal{M}\sim q_{\psi}}
\left[
\left(\sum_{i\in\mathcal{M}} \operatorname{stopgrad}(E_i)\right)
\nabla_{\psi}\log q_{\psi}(\mathcal{M}\mid X)
\right].
\label{eq:reinforce_bridge}
\end{equation}
Eq.~(\ref{eq:aim_loss}) is the practical single-sample surrogate of Eq.~(\ref{eq:reinforce_bridge}) under our fixed-budget Gumbel-Top-K sampling scheme. Hence, the policy is encouraged to reduce the likelihood that high-error patches are selected into the masked set, thereby increasing their tendency to be preserved in the visible set under the fixed-budget sampling scheme. In this sense, AIM does not optimize mutual information in closed form; rather, it optimizes a reconstruction-driven surrogate whose effect is to reduce $H(X\mid X_V)$ and thus increase $I(X_V;X)$.
\subsection{Dual-Stream Convolutional Reconstruction Backbone}
\label{sec:reconstruction}
The core of CIG-MAE is the reconstruction of the original signal from the partial input filtered by AIM. We deliberately employ a pure Convolutional Neural Network (CNN) encoder-decoder architecture. This choice is substantiated by both physical characteristics and practical constraints:

\begin{enumerate}
    \item \textbf{Inductive Bias Match:} CNNs capture local structural and translation-equivariant features, which align with CSI patterns \cite{sensefi_survey}. In contrast, Transformer-based backbones (ViTs) model global dependencies and require massive datasets to learn such local priors \cite{ViT}. However, existing CSI datasets are limited in scale, often ranging from hundreds to thousands of samples \cite{dataset_NLOS_2020,dataset_SignFi_2018,dataset_UTHAR,dataset_wimans}, with only a few reaching tens of thousands \cite{dataset_widar}. This scarcity is insufficient to train data-hungry Transformers effectively.
    \item \textbf{Deployment Efficiency:} WiFi sensing applications are typically deployed on resource-constrained edge devices \cite{survey_edgedevice_limitedMemory}. As empirically validated in Section~\ref{backbone_comparison}, our CNN backbone requires significantly fewer FLOPs and parameters compared to ViTs, making real-time inference feasible without sacrificing accuracy.
\end{enumerate}

\textbf{Reconstruction Process.} 
The masked input $\tilde{\mathbf{X}}^{m}$ is mapped by the encoder $f_{\theta}^{m}$ to a latent representation $\mathbf{z}^{m}$, and subsequently reconstructed by the decoder $g_{\phi}^{m}$ to yield $\hat{\mathbf{X}}^{m}$. To quantify quality, we compute the Mean Absolute Error (MAE) exclusively on the masked pixels $\Omega(\Pi)=\{(n,s,t)\mid \Pi(s,t)=0\}$:
\begin{equation}
\mathcal{L}_{\mathrm{rec}}^{m} = \frac{1}{|\Omega(\Pi)|} \sum_{(n,s,t)\in\Omega(\Pi)} \bigl|\hat{\mathbf{X}}^{m}_{n,s,t}-\mathbf{X}^{m}_{n,s,t}\bigr|.
\end{equation}
The choice of an unnormalized reconstruction target combined with the MAE function is deliberately tailored to the physical properties of CSI. First, regarding the loss metric, we employ MAE over the standard Mean Squared Error (MSE). CSI signals are prone to sporadic, high-amplitude noise spikes due to hardware imperfections and multipath interference. MSE imposes a quadratic penalty on such outliers, which can dominate the gradient and destabilize training. In contrast, MAE's linear penalty provides robustness, ensuring the model focuses on learning the underlying activity structure rather than fitting these transient artifacts. Second, regarding the target, we avoid per-sample normalization during loss computation. Normalization tends to compress the signal's dynamic range, potentially obscuring the semantic distinction between high-magnitude activity bursts and low-magnitude background fluctuations. By preserving the full dynamic range, we force the model to prioritize the reconstruction of significant, activity-induced signal variations. The empirical superiority of this design is validated in Section~\ref{sec:sensitivity}.
\subsection{Cross-Modal Representation Alignment}
\label{sec:alignment}
While the dual streams independently reconstruct their respective modalities, it is crucial that their learned latent representations are semantically consistent. To achieve this without relying on negative samples, we introduce a BT regularizer \cite{SSL_BT_2021}.

To ensure the statistical stability of the correlation estimates, this alignment process is performed on the original, \textit{unmasked} input $\mathbf{X}^{m}$, paralleling the reconstruction path. For each modality $m \in \{A,P\}$, the projection vector $\mathbf{p}^{m}$ is obtained through the cascaded mapping of the encoder $f_{\theta}^{m}$ and the BT projection head $h_{\omega}^{m}$:
\begin{equation}
\mathbf{p}^{m} = h_{\omega}^{m}(f_{\theta}^{m}(\mathbf{X}^{m})).
\end{equation}
Within a data batch of size $B$, we assess the alignment between the two streams by computing the cross-correlation matrix $\mathbf{C}$:
\begin{equation}
\mathbf{C}_{ij} = \frac{1}{B}\sum_{b=1}^{B}\tilde{\mathbf{p}}^{A}_{b,i}\,\tilde{\mathbf{p}}^{P}_{b,j},
\end{equation}
where $\tilde{\mathbf{p}}$ denotes the representations after batch normalization. The BT loss is then defined to strictly penalize deviations from the identity matrix:
\begin{equation}
\mathcal{L}_{\mathrm{BT}} = \sum_{i} (1-\mathbf{C}_{ii})^{2} + \lambda \sum_{i \neq j} \mathbf{C}_{ij}^{2}.
\label{BT_loss}
\end{equation}
The first term (invariance) forces the diagonal elements to 1, aligning the representations of amplitude and phase. The second term (redundancy reduction) forces off-diagonal elements to 0, decorrelating the feature dimensions to maximize information content.
\begin{algorithm}[htbp]
\caption{CIG-MAE Pre-training Procedure}
\label{alg:cig_mae_pretraining}
\begin{algorithmic}[1]
\Require Data $(\mathbf{X}^{A}, \mathbf{X}^{P})$, Backbones $\Theta=\{f_\theta, g_\phi, h_\omega\}$, Policy Nets $\Psi=\{\pi_\psi, h_\eta\}$
\Ensure Optimized Encoders $f_\theta^{A}, f_\theta^{P}$

\While{not converged}
    \State Sample batch $(\mathbf{X}^{A}, \mathbf{X}^{P})$
    
    \State \textit{// Phase 1: Adaptive Masking (Sec. \ref{sec:aim})}
    \State // Policy network $\pi_\psi$ estimates importance; $\mathcal{M}$ is sampled via Gumbel-Top-K
    \State $\mathcal{M}^{A} \leftarrow \text{AIM}(\mathbf{X}^{A}, \Psi^{A})$; $\mathcal{M}^{P} \leftarrow \text{AIM}(\mathbf{X}^{P}, \Psi^{P})$
    
    \State \textit{// Phase 2: Dual-Stream Reconstruction (Sec. \ref{sec:reconstruction})}
    \State // Apply mask and reconstruct using Encoder $f_\theta$ and Decoder $g_\phi$
    \State $\tilde{\mathbf{X}}^{m} \leftarrow \mathbf{X}^{m} \odot (\mathbf{1}-\mathcal{M}^{m})$ for $m \in \{A, P\}$ 
    \State $\hat{\mathbf{X}}^{m} \leftarrow g_\phi^m(f_\theta^m(\tilde{\mathbf{X}}^{m}))$ for $m \in \{A, P\}$ 
    \State $\mathcal{L}_{\text{rec}} \leftarrow \text{MaskedMAE}(\hat{\mathbf{X}}^{A,P}, \mathbf{X}^{A,P}, \mathcal{M}^{A,P})$ 

    \State \textit{// Phase 3: Cross-Modal Alignment (Sec. \ref{sec:alignment})}
    \State // Project unmasked features via $h_\omega$ for Barlow Twins
    \State $\mathbf{p}^{A} \leftarrow h_\omega^A(f_\theta^A(\mathbf{X}^A))$; $\mathbf{p}^{P} \leftarrow h_\omega^P(f_\theta^P(\mathbf{X}^P))$ 
    \State $\mathcal{L}_{\text{BT}} \leftarrow \text{BTLoss}(\mathbf{p}^{A}, \mathbf{p}^{P})$ 

    \State \textit{// Phase 4: Decoupled Parameter Optimization}
    \State // 1. Update Policy $\Psi$: maximize reconstruction error of masked patches
    \State $E^{A,P} \leftarrow \text{PixelWiseError}(\hat{\mathbf{X}}^{A,P}, \mathbf{X}^{A,P})$ 
    \State $\mathcal{L}_{\text{AIM}} \leftarrow \text{PolicyGradient}(E^{A,P}, \mathcal{M}^{A,P}, \Psi)$ 
    \State $\Psi \leftarrow \text{Optimizer}(\nabla_{\Psi} (w_{\text{aim}}\mathcal{L}_{\text{AIM}}))$
    
    \State // 2. Update Backbone $\Theta$: minimize reconstruction and redundancy
    \State $\Theta \leftarrow \text{Optimizer}(\nabla_{\Theta} (\mathcal{L}_{\text{rec}} + w_{\text{bt}}\mathcal{L}_{\text{BT}}))$
\EndWhile

\State \Return Optimized encoders $f_\theta^{A}, f_\theta^{P}$
\end{algorithmic}
\end{algorithm}
\subsection{Overall Training Objective}
The total loss is a weighted sum of the three objectives:
\begin{equation}
\mathcal{L}_{\text{CIG-MAE}} = \mathcal{L}_{\text{rec}} + w_{\text{aim}}\mathcal{L}_{\text{AIM}} + w_{\text{bt}}\mathcal{L}_{\text{BT}}.
\end{equation}
We employ a \textbf{decoupled update strategy} to distinguish the optimization targets of the reconstruction and masking-policy components. Specifically, the backbone parameters $\Theta = \{f_\theta, g_\phi, h_\omega\}$ are updated by minimizing the reconstruction and alignment losses ($\mathcal{L}_{\text{rec}} + w_{\text{bt}}\mathcal{L}_{\text{BT}}$), while the policy parameters $\Psi = \{\pi_\psi, h_\eta\}$ are updated solely by the policy gradient loss $\mathcal{L}_{\text{AIM}}$. This separation ensures that the policy network independently learns to identify information-dense regions without interference from the reconstruction objective. The complete pre-training procedure is summarized in Algorithm~\ref{alg:cig_mae_pretraining}.
\begin{table*}[t]
\centering
\caption{Network architecture used in CIG-MAE. E, D, and F represent Encoder, Decoder, and Classifier, respectively.}
\label{Arch_exp}
\resizebox{\textwidth}{!}{
\begin{tabular}{c|c|c|c}
\toprule
\textbf{Layer} & \textbf{$E$ (Encoder)} & \textbf{$D$ (Decoder)} & \textbf{$F$ (Classifier)} \\
\midrule
\multicolumn{4}{c}{Input CSI: $3 (6)\times30\times200$ (antenna $\times$ subcarrier $\times$ timestamp)} \\
\midrule
1 & Conv, $C=128$, $K=(3,5)$, $S=(3,5)$ & MLP: $256\!\to\!512\times5\times10$ & FC: $512\!\to\!C$, Softmax\\
\hline
2 & Conv, $C=256$, $K=(2,2)$, $S=(2,2)$ & DeConv, $C=512$, $K=(1,2)$, $S=(1,2)$ & --- \\
\hline
3 & Conv, $C=512$, $K=(1,2)$, $S=(1,2)$ & DeConv, $C=256$, $K=(2,2)$, $S=(2,2)$ & --- \\
\hline
4 & MLP: $512\times5\times10\!\to\!256$ & DeConv, $C=128$, $K=(3,5)$, $S=(3,5)$ & --- \\
\bottomrule
\end{tabular}
}
\end{table*}
\section{Experiments}
\label{sec:experiments}
This section systematically evaluates the proposed CIG\mbox{-}MAE framework. First, we validate its effectiveness by comparing it with several SOTA self-supervised learning methods on three public CSI datasets. Subsequently, the model's behavior and performance are analyzed through ablation studies, hyperparameter sensitivity analysis, and settings with varying amounts of labeled data.
\subsection{Experimental Setup}
\label{Experimental_Setup}
\textit{Datasets.} We evaluate CIG-MAE on three datasets representing diverse sensing scenarios:
\begin{itemize}
    \item \textbf{SignFi} \cite{dataset_SignFi_2018} (Sign Language): A subset from User 5 (home environment), comprising 2,760 samples across 276 classes, collected using an 802.11n CSI tool \cite{80211CSItools}. The input tensor shape is $3\times 30\times 200$.
    \item \textbf{NLOS} \cite{dataset_NLOS_2020} (Daily Activities): Data from Environment 1, consolidated into 6 classes (3,000 samples) following \cite{NLOS_class_1,NLOS_class_2}. The subjects in Environment 1 consist of 10 males with distinct ages, weights, and heights. Variable durations $[754, 1601]$ were downsampled to 200.
    \item \textbf{HTHI} \cite{dataset_HTHI_2020} (Human-Interaction): Contains 4,800 trials of 12 interaction types from 40 pairs. The 40 pairs were randomly formed from a pool of 66 subjects with varying ages, weights, and heights. Durations $[1040, 2249]$ were downsampled to 200.
\end{itemize}

\textit{Protocol \& Implementation.} 
We implement the model in PyTorch on an NVIDIA RTX 5090. Datasets are stratified (8:2 train/test split), so the training splits of SignFi, NLOS and HTHI contain $2{,}208$, $2{,}400$ and $3{,}840$ unlabeled sequences, respectively. Raw CSI phase is linearly calibrated \cite{Preprocess_phaseSanitation_2012}, and both modalities are z-score normalized. 
We adhere to the standard $k$-shot linear probing protocol: pre-trained encoders are frozen, and concatenated representations are fed to a linear classifier. 
\textbf{Pre-training} runs for 300 epochs (batch size 256) using AdamW ($lr=1\times 10^{-4}$, weight decay $0.01$, $\beta=(0.9,0.95)$). We set the mask ratio $\rho=0.95$, patch size $(3,5)$, and loss weights $w_{\text{bt}}=0.2, w_{\text{aim}}=1\times 10^{-4}$. 
The AIM policy features $\mathbf{x}^{m}$ and encoder latent features $\mathbf{z}^{m}$ are both 256. Critically, the BT projection head is a 3-layer MLP with width 1024 (SignFi/HTHI) or 5096 (NLOS). \textbf{Downstream} evaluation uses 1-shot (SignFi) or 10-shot (NLOS/HTHI) learning for 100 epochs (batch size 32, $lr=1\times 10^{-3}$). Architecture details are provided in Table I. We adopt this mid-sized dual-stream CNN backbone to practically balance representation expressiveness and deployment efficiency. Smaller variants lack sufficient receptive field and channel capacity to capture the spatio-temporal structure of $3\times 30\times 200$ CSI maps, whereas larger models incur prohibitive latency and memory overheads on edge devices with diminishing empirical gains.

\subsection{Baselines}
We compare CIG-MAE against a fully supervised upper bound and three SOTA self-supervised methods:
\begin{itemize}
    \item \textbf{Supervised Model}: Trained end-to-end on the full labeled dataset using the exact same backbone as CIG-MAE, serving as the performance upper bound.
    \item \textbf{AutoFi} \cite{SSL_AutoFi_Wifi_2023}: A contrastive method that jointly optimizes probabilistic consistency, mutual information maximization, and geometric structure consistency.
    \item \textbf{AutoSen} \cite{SSL_Autosen_Wifi_2024}: A cross-modal autoencoder that learns joint representations through the asymmetric proxy task of reconstructing phase from amplitude.
    \item \textbf{CAPC} \cite{SSL_CAPC_2024}: Combines CPC with BT. It predicts future latent representations to capture temporal context while using BT to reduce feature redundancy.
\end{itemize}

\subsection{Performance Comparison}
Table~\ref{results_compare} compares CIG-MAE against SOTA SSL methods and a supervised baseline. CIG-MAE consistently outperforms all SSL counterparts. Notably, on SignFi, it achieves 98.49\% accuracy, surpassing even the supervised baseline (98.19\%), indicating superior generalization.
This advantage widens in complex scenarios (NLOS and HTHI), where CIG-MAE maintains robustness with accuracies of 59.80\% and 38.90\% respectively, establishing a clear margin over all SSL baselines.
\begin{table}[htbp]
\centering
\caption{Performance (Mean $\pm$ Std) comparison of different methods on SignFi, NLOS, and HTHI datasets with a linear classifier.}
\label{results_compare}
\small
\resizebox{\columnwidth}{!}{
\begin{tabular}{l c c c c c c}
\toprule
\multirow{2}{*}{Method} & \multicolumn{2}{c}{SignFi} & \multicolumn{2}{c}{NLOS} & \multicolumn{2}{c}{HTHI} \\
\cmidrule(lr){2-3} \cmidrule(lr){4-5} \cmidrule(lr){6-7}
 & {Accuracy} & {F1 Score} & {Accuracy} & {F1 Score} & {Accuracy} & {F1 Score} \\
\midrule
Supervised & $98.19 \pm 2.56$ & $97.88 \pm 3.01$ & $99.66 \pm 0.23$ & $99.67 \pm 0.27$ & $97.76 \pm 0.37$ & $97.76 \pm 0.37$ \\
AutoFi     & $94.60 \pm 0.21$ & $94.41 \pm 0.06$ & $48.59 \pm 0.83$ & $47.73 \pm 1.49$ & $30.99 \pm 1.54$ & $29.88 \pm 2.20$ \\
AutoSen    & $88.59 \pm 1.28$ & $87.85 \pm 1.27$ & $48.08 \pm 5.30$ & $48.19 \pm 3.56$ & $30.95 \pm 2.33$ & $30.10 \pm 1.63$ \\
CAPC       & $95.56 \pm 1.92$ & $95.63 \pm 1.46$ & $46.91 \pm 0.82$ & $46.02 \pm 2.21$ & $34.06 \pm 2.35$ & $35.05 \pm 1.81$ \\
CIG-MAE    & $98.49 \pm 0.27$ & $98.42 \pm 0.20$ & $59.80 \pm 3.48$ & $57.93 \pm 3.30$ & $38.90 \pm 1.93$ & $39.43 \pm 1.57$ \\
\bottomrule
\end{tabular}
}
\end{table}

The performance gap stems from the misalignment of baselines with CSI characteristics:
\begin{itemize}
    \item \textbf{AutoFi} falters in multi-user settings (NLOS, HTHI) as its contrastive paradigm relies on simplistic augmentations insufficient to handle high intra-class variance.
    \item \textbf{CAPC} imposes a restrictive temporal continuity bias, causing it to neglect self-contained spatio-temporal patterns critical for fine-grained actions.
    \item \textbf{AutoSen}'s asymmetric design creates an information bottleneck by assuming amplitude fully specifies phase, thereby losing unique phase-dependent information.
\end{itemize}

In contrast, CIG-MAE overcomes these limitations through a synergistic design. Its masked autoencoding paradigm captures holistic intrinsic structures rather than brittle inter-sample relations. The symmetric, BT-regularized architecture ensures complete, decorrelated fusion of amplitude and phase. Crucially, AIM dynamically focuses the learning budget on discriminative regions, ensuring robustness against noise where other models fail.

\subsection{Ablation Study}
We validate the contribution of each component by systematically dismantling the full model (Table \ref{Ablation}).

\begin{table}[htbp]
\centering
\caption{The performance (Mean $\pm$ Std) with different design choices for SignFi, NLOS, and HTHI data.}
\label{Ablation}
\resizebox{\columnwidth}{!}{
\begin{tabular}{l c c c c c c}
\toprule
\multirow{2}{*}{Method} & \multicolumn{2}{c}{SignFi} & \multicolumn{2}{c}{NLOS} & \multicolumn{2}{c}{HTHI} \\
\cmidrule(lr){2-3} \cmidrule(lr){4-5} \cmidrule(lr){6-7}
 & {Accuracy} & {F1 Score} & {Accuracy} & {F1 Score} & {Accuracy} & {F1 Score} \\
\midrule
Single-Stream (Amp.) & 96.43$\pm$0.73 & 96.20$\pm$0.67 & 56.53$\pm$3.40 & 54.35$\pm$3.24 & 37.03$\pm$0.73 & 37.90$\pm$0.83 \\
Dual-Stream MAE & 97.48$\pm$0.97 & 97.35$\pm$1.03 & 52.99$\pm$4.07 & 53.78$\pm$3.04 & 37.03$\pm$0.52 & 37.74$\pm$0.47 \\
w/o BT & 97.88$\pm$0.73 & 97.56$\pm$0.82 & 52.33$\pm$3.58 & 54.05$\pm$3.16 & 36.82$\pm$1.02 & 37.30$\pm$0.92 \\
w/o AIM & 97.70$\pm$0.69 & 97.45$\pm$0.66 & 57.50$\pm$3.64 & 57.33$\pm$3.85 & 37.29$\pm$2.10 & 38.18$\pm$2.30 \\
CIG-MAE & \textbf{98.49$\pm$0.27} & \textbf{98.42$\pm$0.20} & \textbf{59.80$\pm$3.48} & \textbf{57.93$\pm$3.30} & \textbf{38.90$\pm$1.93} & \textbf{39.43$\pm$1.57} \\
\bottomrule
\end{tabular}
}
\end{table}

\begin{figure}[h]
\centering
\subfigure[AIM-guided Amplitude Masking\label{overlay_heat_amp_aim}]
{\includegraphics[width=0.48\columnwidth]{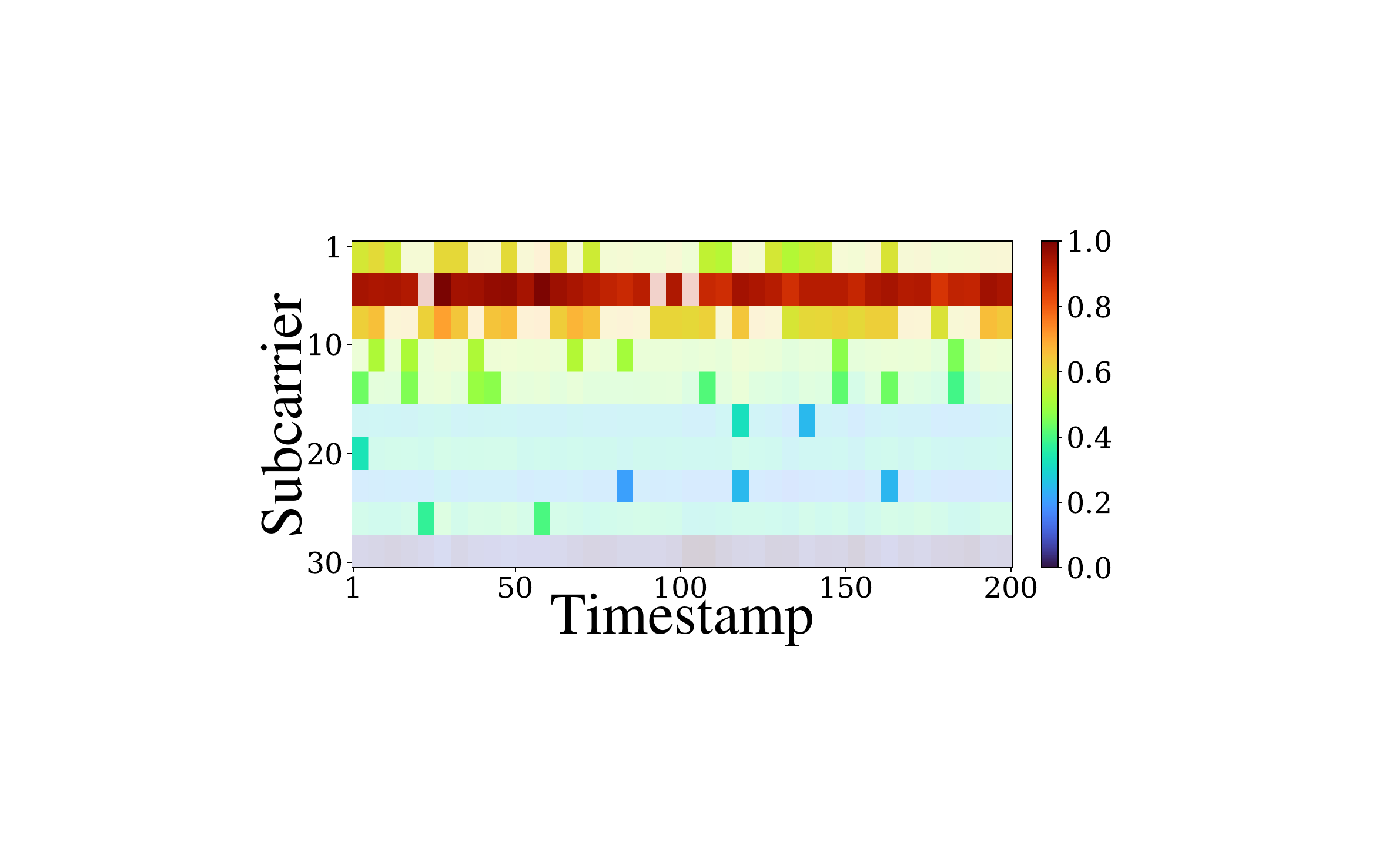}}\hspace{2mm}
\subfigure[AIM-guided Phase Masking\label{overlay_heat_phase_aim}]
{\includegraphics[width=0.48\columnwidth]{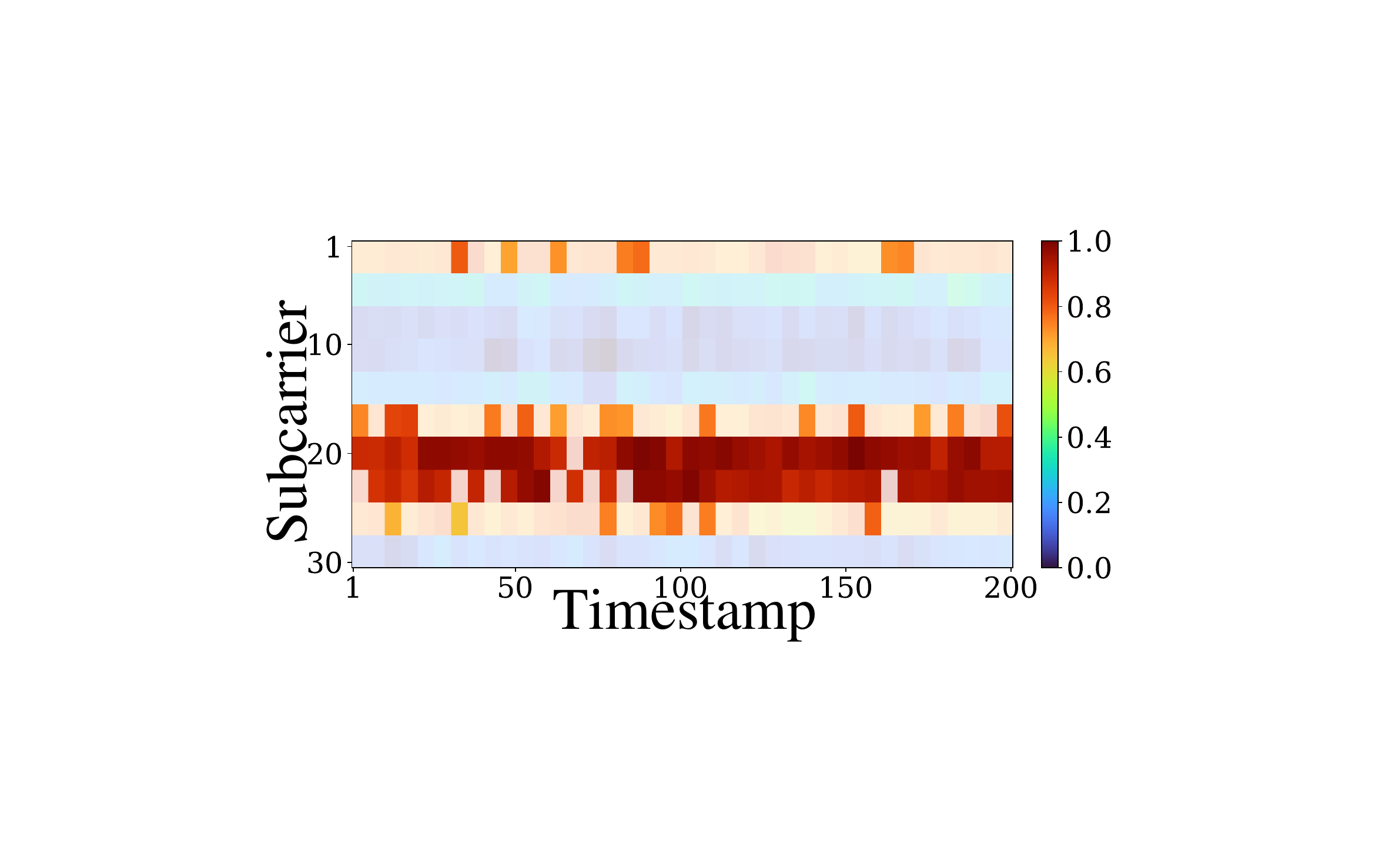}}

\vspace{1mm}

\subfigure[Random Amplitude Masking\label{overlay_heat_amp_random}]
{\includegraphics[width=0.48\columnwidth]{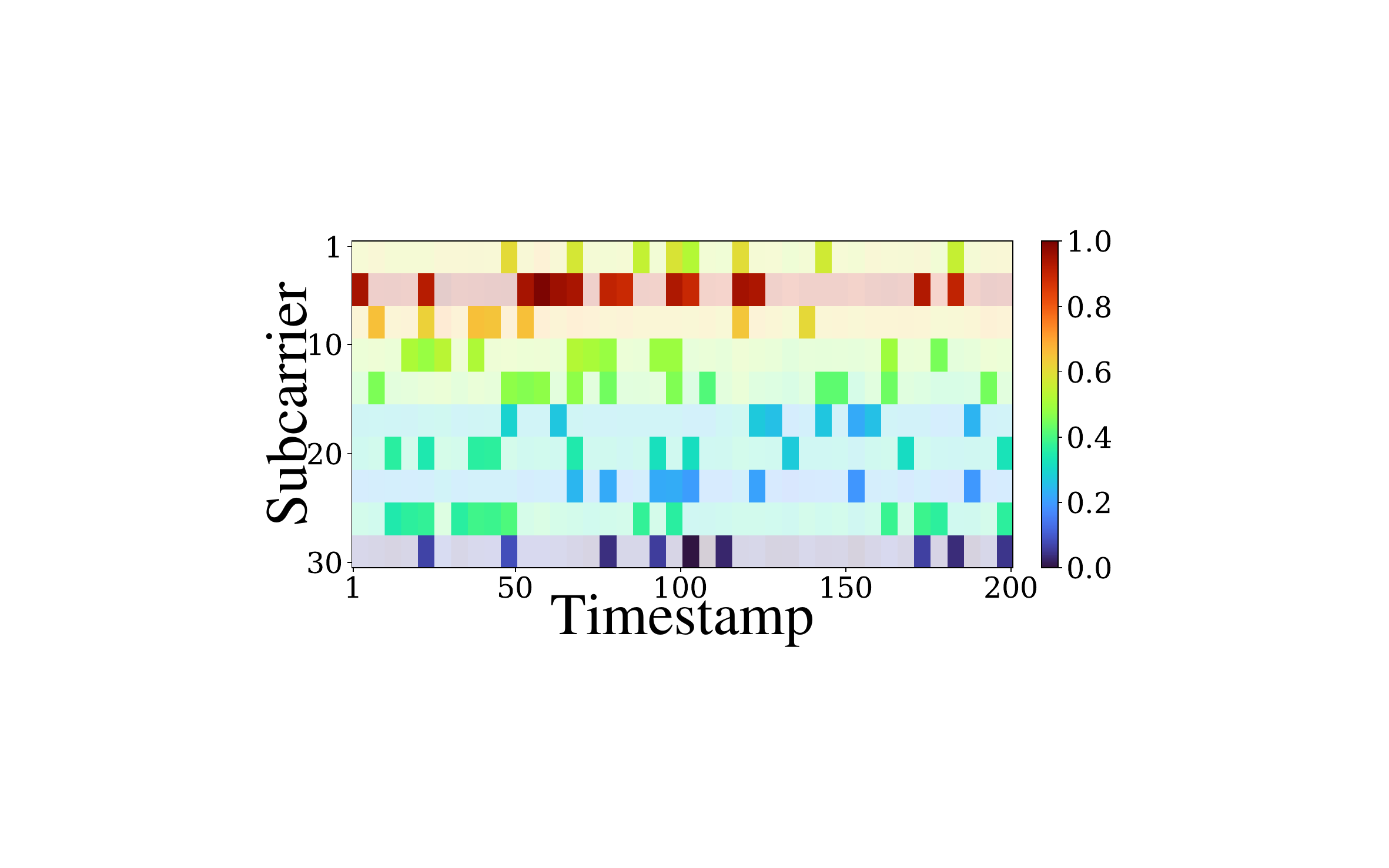}}\hspace{2mm}
\subfigure[Random Phase Masking\label{overlay_heat_phase_random}]
{\includegraphics[width=0.48\columnwidth]{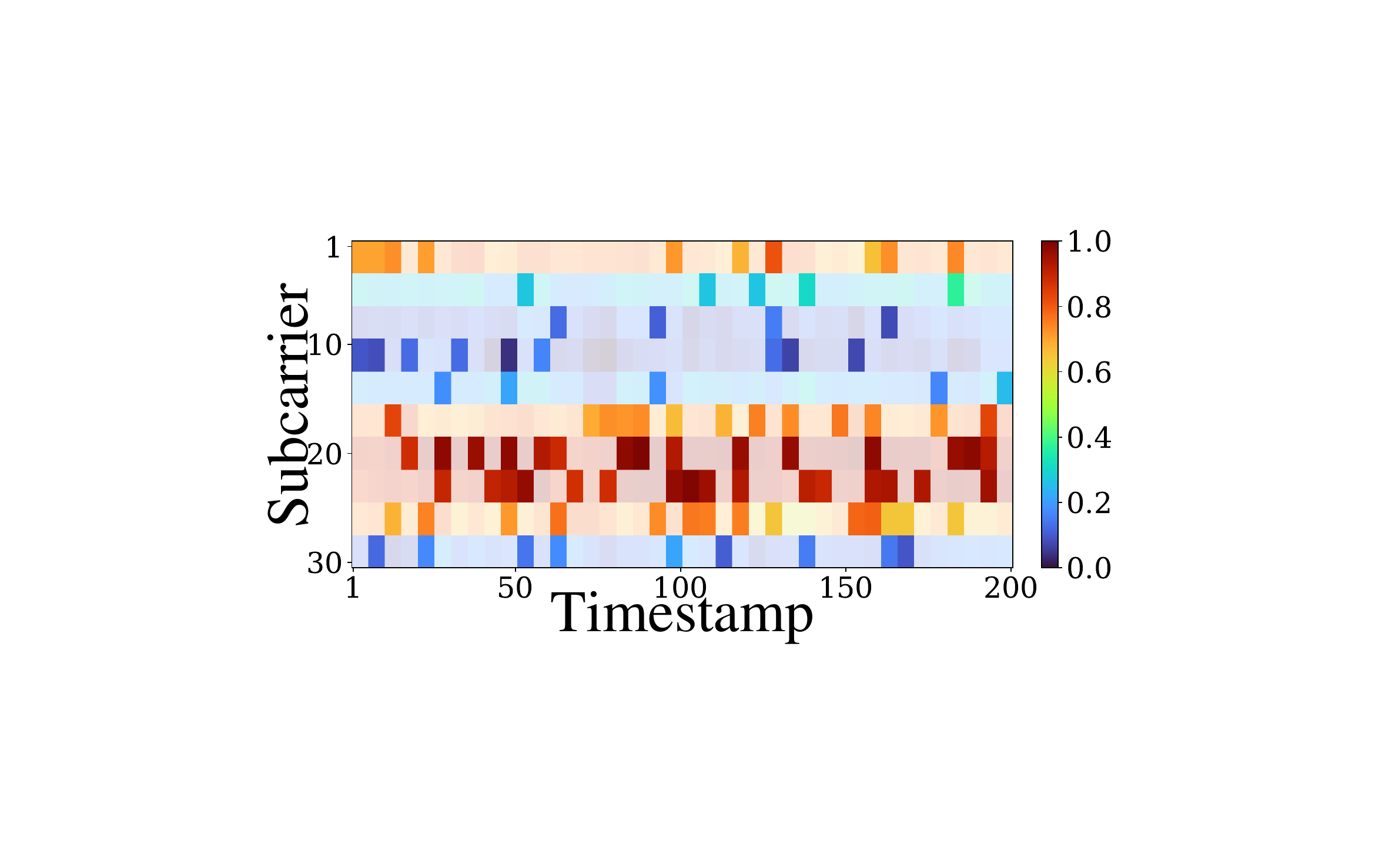}}

\caption{Demonstration of AIM's visibility allocation. Compared to random masking, AIM suppresses low-information regions and increases the visibility of high-information regions, concentrating the learning budget on critical time-frequency blocks related to the activity, thereby obtaining more discriminative representations.}
\label{AIM_analysis_viz}
\end{figure}

\textbf{Impact of AIM.} Replacing AIM with random masking (\textit{w/o AIM}) leads to a consistent performance degradation across all datasets. Specifically, removing the AIM strategy results in an accuracy drop of 2.30\% on NLOS (from 59.80\% to 57.50\%) and 1.61\% on HTHI (from 38.90\% to 37.29\%). This confirms that in low-SNR and complex interaction scenarios, strictly random masking is insufficient. By focusing on information-dense regions, AIM enables the model to learn more discriminative features from limited visibility, as visualized in Fig.~\ref{AIM_analysis_viz}.

\begin{figure}[htbp]
\centering
\subfigure[Amplitude-Phase Correlation Matrix without BT Constraint\label{NoBT}]
{\includegraphics[width=0.48\columnwidth]{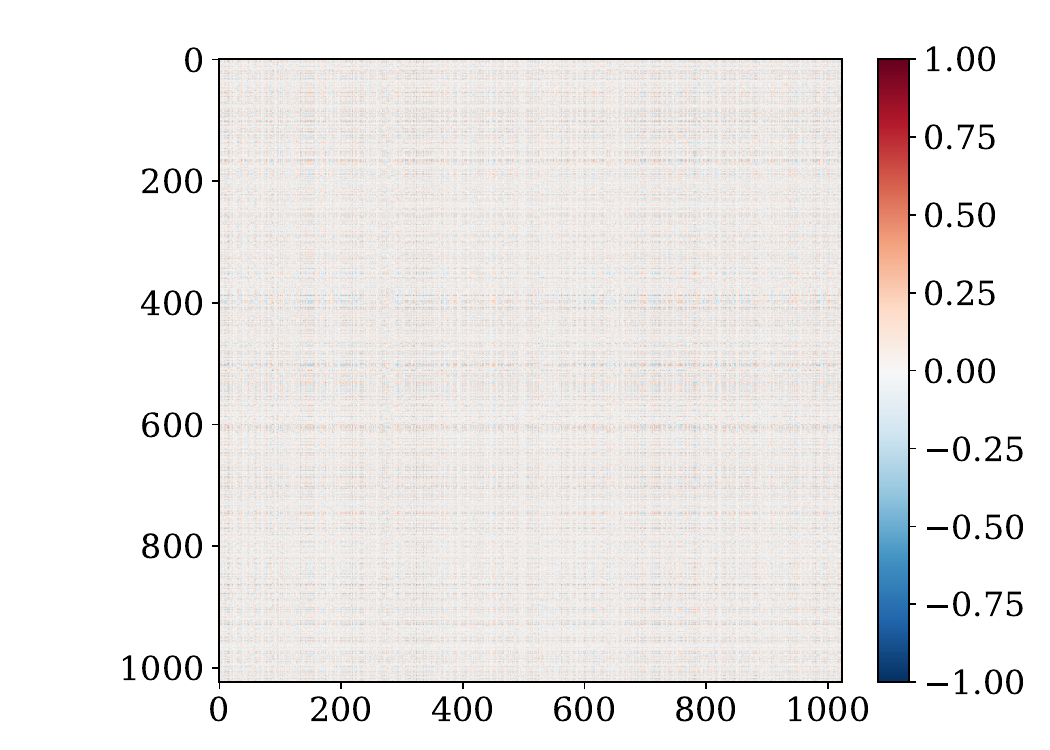}}\hspace{2mm}
\subfigure[Amplitude-Phase Correlation Matrix with BT Constraint\label{WithBT}]
{\includegraphics[width=0.48\columnwidth]{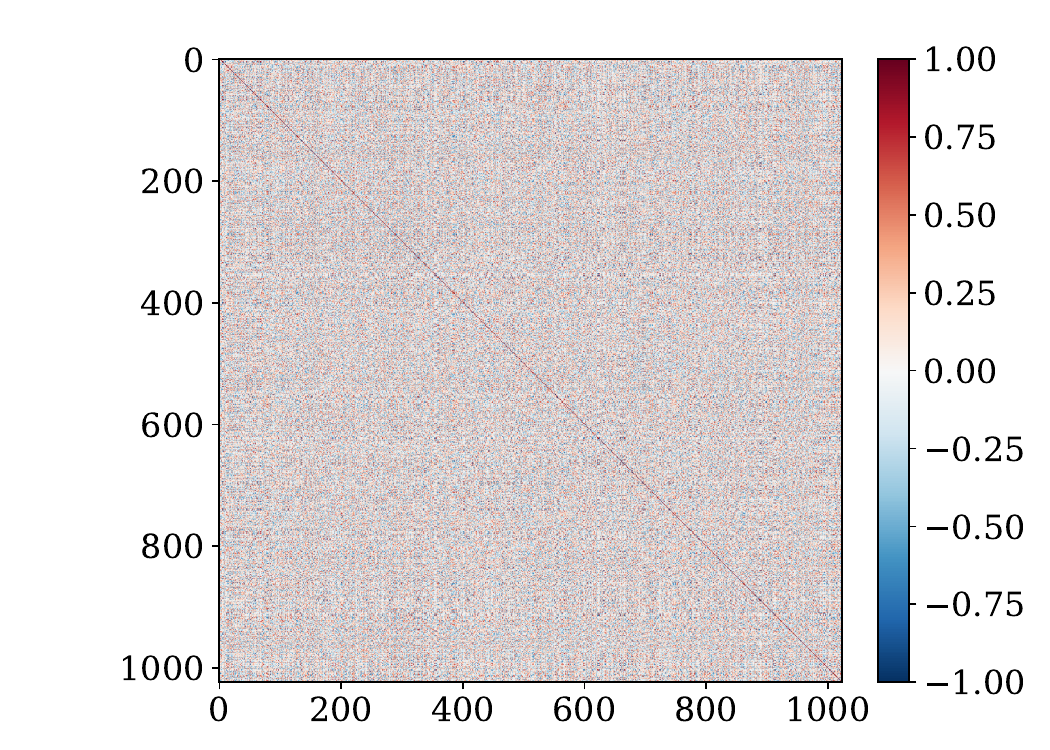}}
\caption{Demonstration of the alignment and decorrelation effect of BT. After introducing BT, the diagonal elements of the correlation matrix approach 1, and the off-diagonal elements approach 0, reflecting enhanced consistency and reduced redundancy in the corresponding dimensions of the two streams.}
\label{bt_compare}
\end{figure}

\textbf{Impact of BT Regularizer.} BT proves essential for coordinating the dual streams, especially in noisy environments. Removing it (\textit{w/o BT}) causes a significant 7.47\% accuracy drop on NLOS. Notably, on the challenging NLOS dataset, \textit{w/o BT} performs even worse than the baseline \textit{Dual-Stream MAE}. This suggests that without the alignment constraint provided by BT, the AIM-guided dual streams may diverge towards modality-specific artifacts or noise, failing to learn a coherent representation. Fig.~\ref{bt_compare} demonstrates how BT effectively aligns representations and reduces redundancy.

\textbf{Dual-Stream Architecture.} Comparison between \textit{Single-Stream} and \textit{Dual-Stream MAE} reveals that naively combining amplitude and phase improves performance on SignFi but degrades it on NLOS. This conflict underscores the necessity of the BT regularizer to resolve feature redundancy, validating the synergistic design of the full CIG-MAE model.

\subsection{Sensitivity Analysis}
\label{sec:sensitivity}
\begin{figure}[htbp]
\centering
% First Row: Dimension-related Parameters
\subfigure[AIM Feature Dimension ($\mathbf{x}^{m}$)\label{fig:sens_aim}]
{\includegraphics[width=0.48\columnwidth]{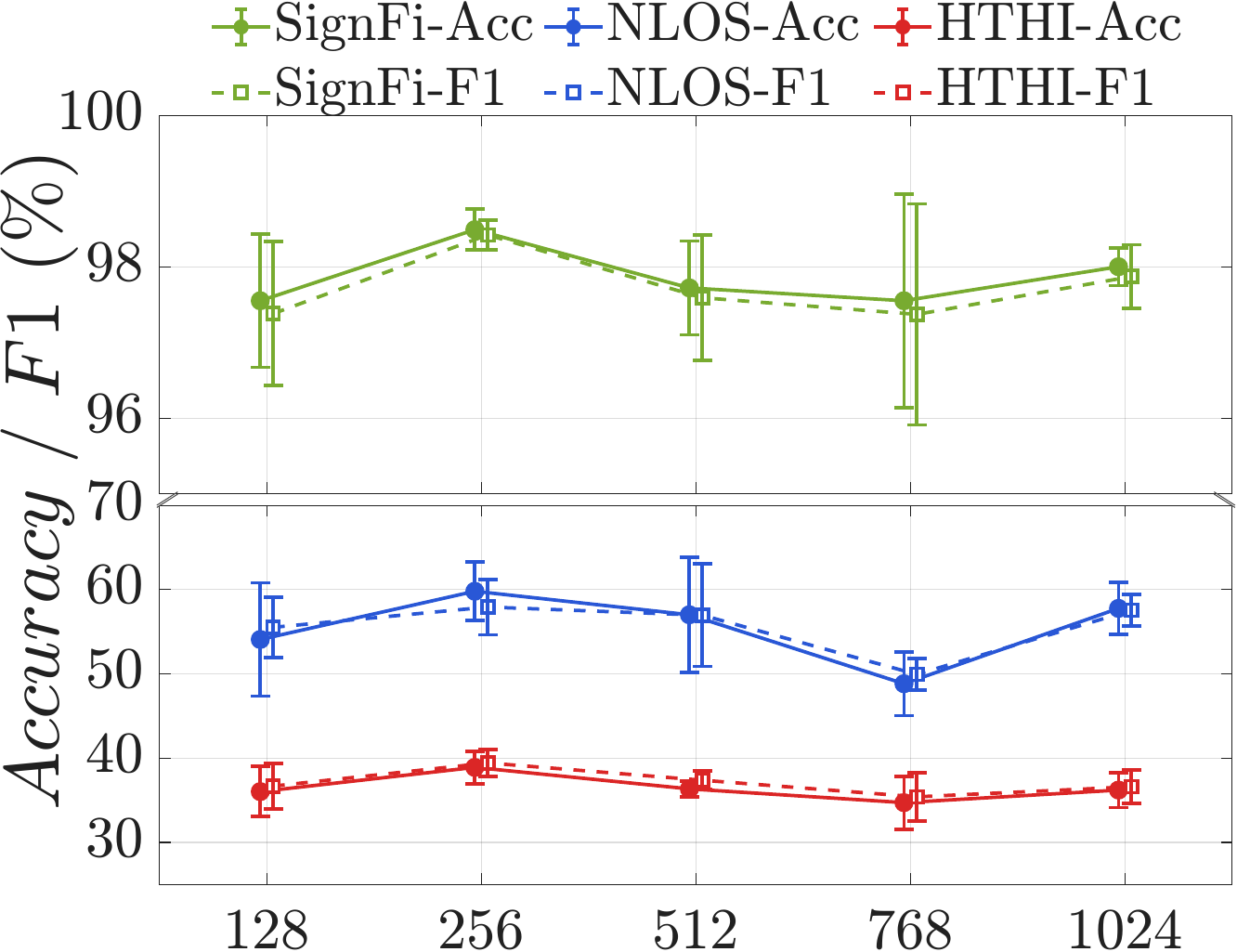}}\hspace{2mm}
\subfigure[Projection Dimension ($\omega^{m}$)\label{fig:sens_proj}]
{\includegraphics[width=0.48\columnwidth]{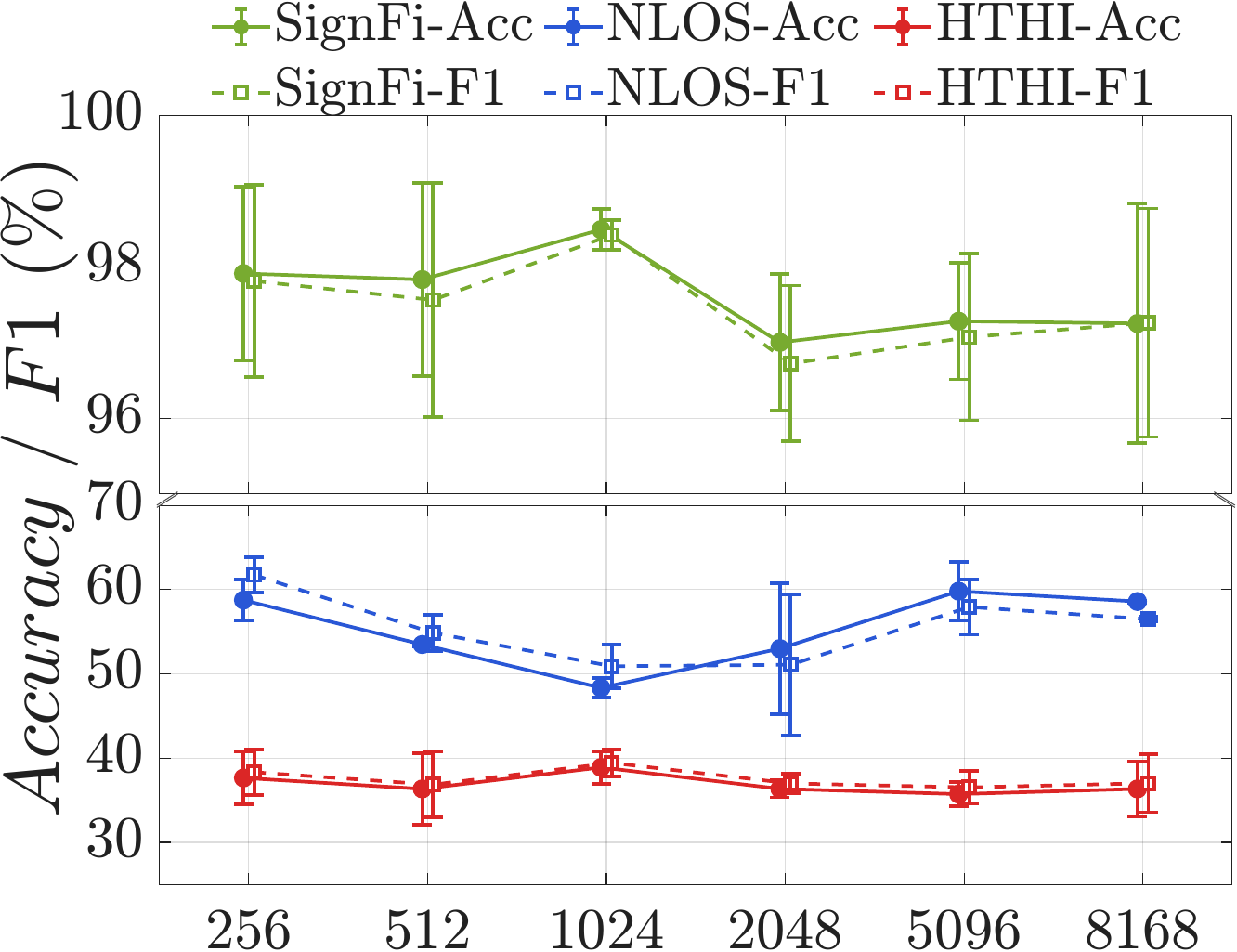}}
\vspace{0mm} % vertical spacing between the rows
% Second Row: Training Objective Parameters
\subfigure[BT Loss Weight ($w_{\text{bt}}$)\label{fig:sens_bt}]
{\includegraphics[width=0.48\columnwidth]{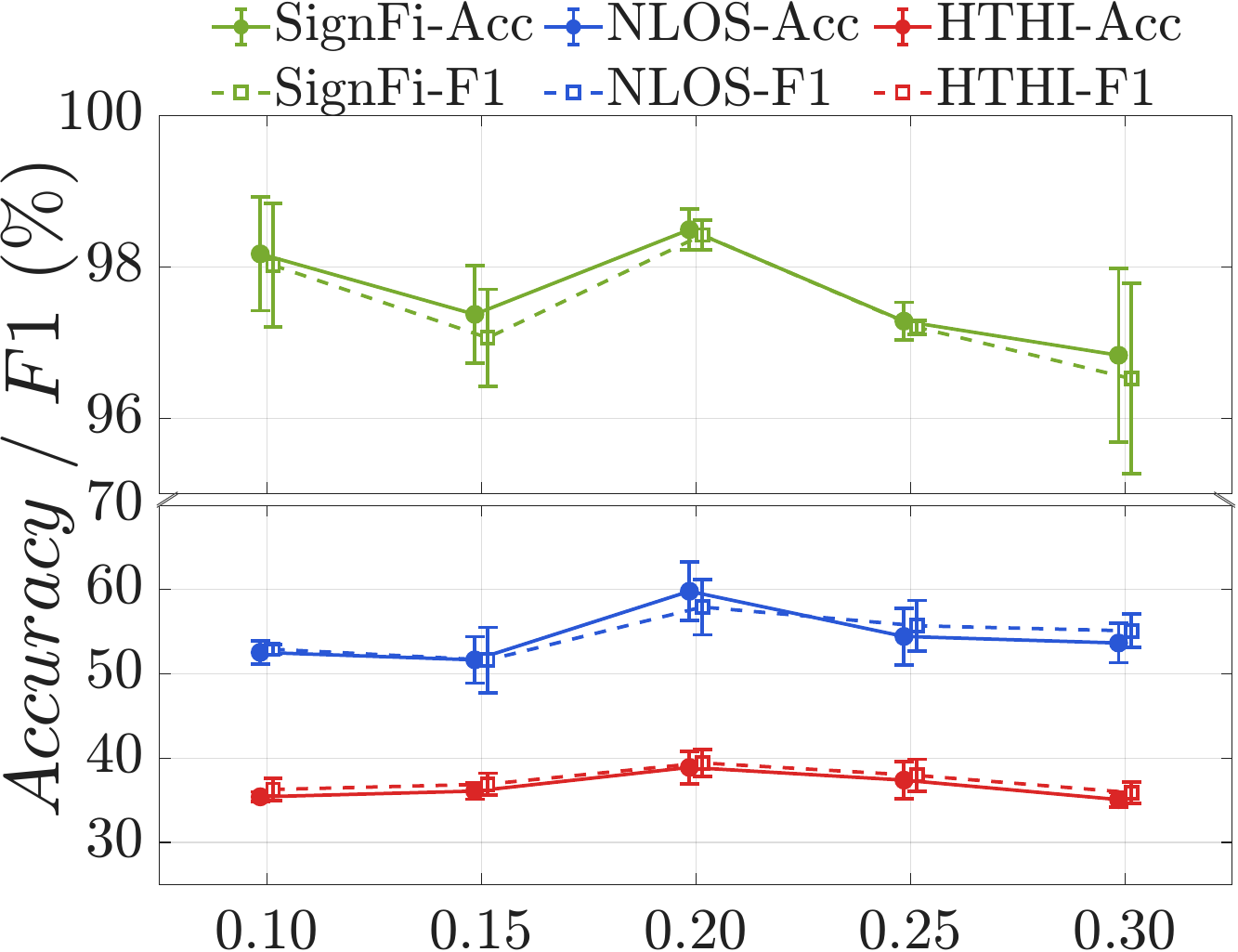}}\hspace{2mm}
\subfigure[Mask Ratio ($\rho$)\label{fig:sens_mask}]
{\includegraphics[width=0.48\columnwidth]{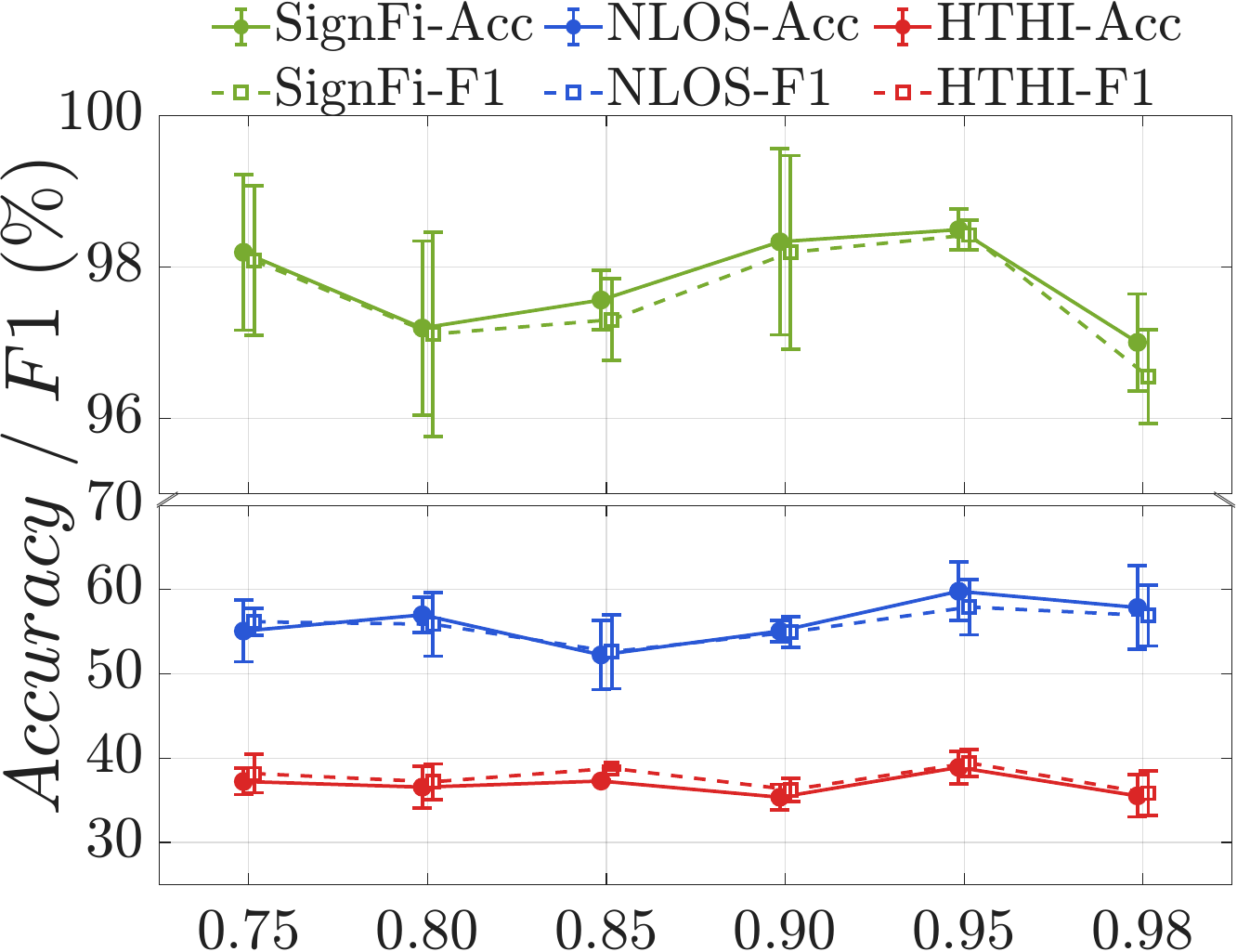}}
\caption{Sensitivity analysis of CIG-MAE's key hyperparameters across the SignFi, NLOS, and HTHI datasets. The results show that while some parameters are stable, others adapt to data characteristics, demonstrating the model's robustness and flexibility.}
\label{fig:sensitivity_analysis}
\end{figure}

We analyze key hyperparameters in Fig.~\ref{fig:sensitivity_analysis} and reconstruction objectives in Table~\ref{loss_function}.

\textbf{Hyperparameters.} The AIM feature dimension ($\mathbf{x}^{m}$) consistently peaks at 256; lower dimensions lack expressive capacity, while higher ones risk overfitting to spurious correlations. The BT loss weight ($w_{\text{bt}}$) stabilizes at 0.2, striking a critical balance: lower weights provide insufficient regularization, while higher weights over-optimize decorrelation at the expense of preserving fine-grained signal details.
In contrast, the projection head dimension ($\omega^{m}$) adapts to dataset complexity. While 1024 suffices for SignFi/HTHI, the noisier NLOS dataset requires a larger capacity (5096) to effectively disentangle features.
Finally, a high mask ratio ($\rho=0.95$) is consistently optimal, confirming that forcing reconstruction from sparse inputs (5\%) compels the model to learn deep intrinsic structures rather than memorizing superficial patterns.

\begin{table}[htbp]
\centering
\caption{Impact of different reconstruction loss functions on CIG-MAE performance (Mean $\pm$ Std) for SignFi, NLOS, and HTHI data.}
\label{loss_function}
\small
\resizebox{\columnwidth}{!}{
\begin{tabular}{l c c c c c c}
\toprule
\multirow{2}{*}{Method} & \multicolumn{2}{c}{SignFi} & \multicolumn{2}{c}{NLOS} & \multicolumn{2}{c}{HTHI} \\
\cmidrule(lr){2-3} \cmidrule(lr){4-5} \cmidrule(lr){6-7}
 & {Accuracy} & {F1 Score} & {Accuracy} & {F1 Score} & {Accuracy} & {F1 Score} \\
\midrule
MAE (w/o Norm.) & \textbf{98.49$\pm$0.27} & \textbf{98.42$\pm$0.20} & \textbf{59.80$\pm$3.48} & \textbf{57.93$\pm$3.30} & \textbf{38.90$\pm$1.93} & \textbf{39.43$\pm$1.57} \\
MAE (w/ Norm.) & 97.09$\pm$1.03 & 96.98$\pm$0.97 & 54.39$\pm$1.44 & 55.43$\pm$1.38 & 35.97$\pm$0.77 & 36.32$\pm$1.11 \\
MSE (w/o Norm.) & 95.82$\pm$1.03 & 95.63$\pm$1.25 & 51.83$\pm$5.57 & 52.13$\pm$3.51 & 37.08$\pm$1.29 & 37.92$\pm$1.53 \\
MSE (w/ Norm.) & 95.38$\pm$0.38 & 95.15$\pm$0.56 & 55.33$\pm$2.08 & 54.46$\pm$1.96 & 36.15$\pm$1.59 & 36.97$\pm$1.47 \\
\bottomrule
\end{tabular}
}
\end{table}

\textbf{Loss Function.} 
Table~\ref{loss_function} shows that \textit{MAE (w/o Norm.)} consistently yields the best performance. MAE is preferred over MSE because its linear penalty is robust to sporadic high-amplitude noise spikes (outliers), avoiding the gradient dominance caused by quadratic penalties. Furthermore, omitting normalization preserves the original dynamic range, which is crucial for distinguishing high-magnitude activity bursts from low-magnitude background fluctuations.

\subsection{Data Efficiency Analysis}
\begin{figure}[htbp]
\centering
\subfigure[SignFi (Accuracy)\label{fig:ratio_signfi_acc}]
{\includegraphics[width=0.48\columnwidth]{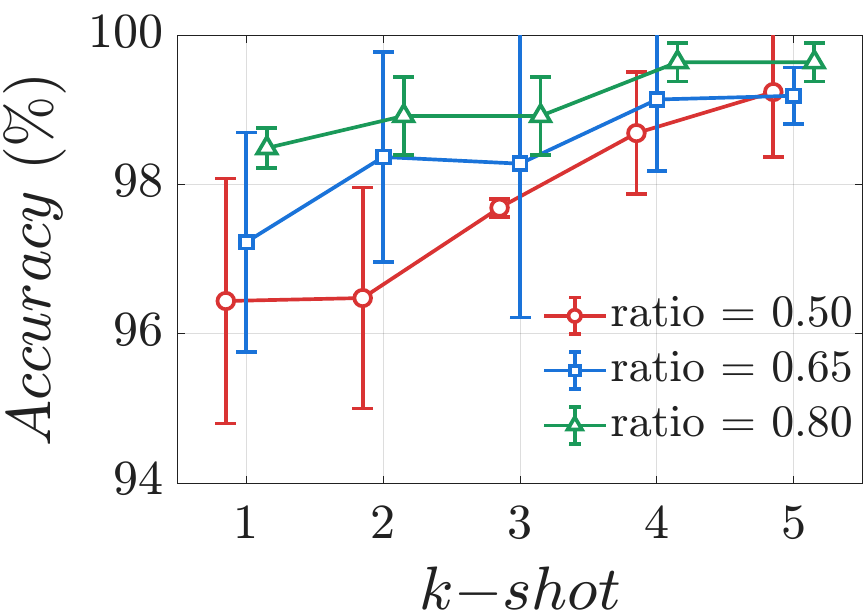}}\hspace{2mm}
\subfigure[SignFi (F1 Score)\label{fig:ratio_signfi_f1}]
{\includegraphics[width=0.48\columnwidth]{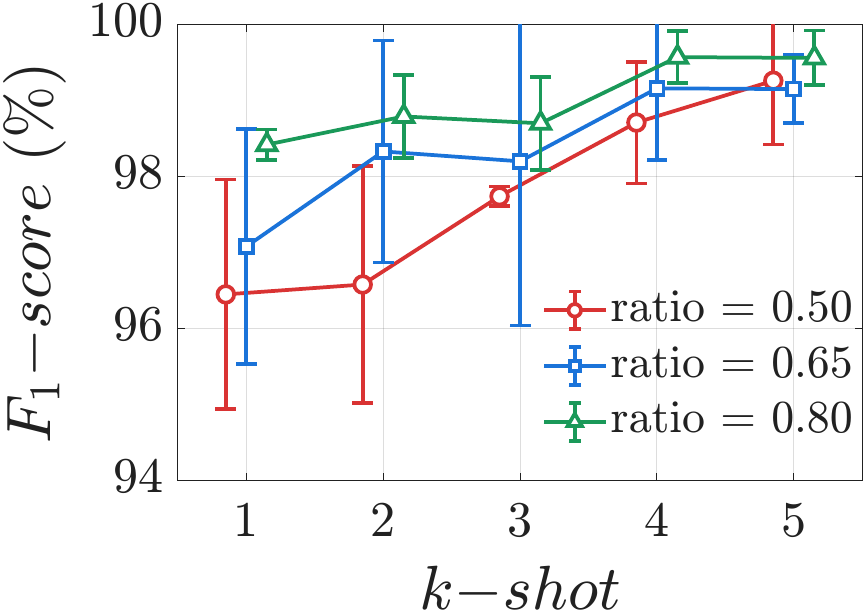}}
\vspace{1mm}
\subfigure[NLOS (Accuracy)\label{fig:ratio_me_acc}]
{\includegraphics[width=0.48\columnwidth]{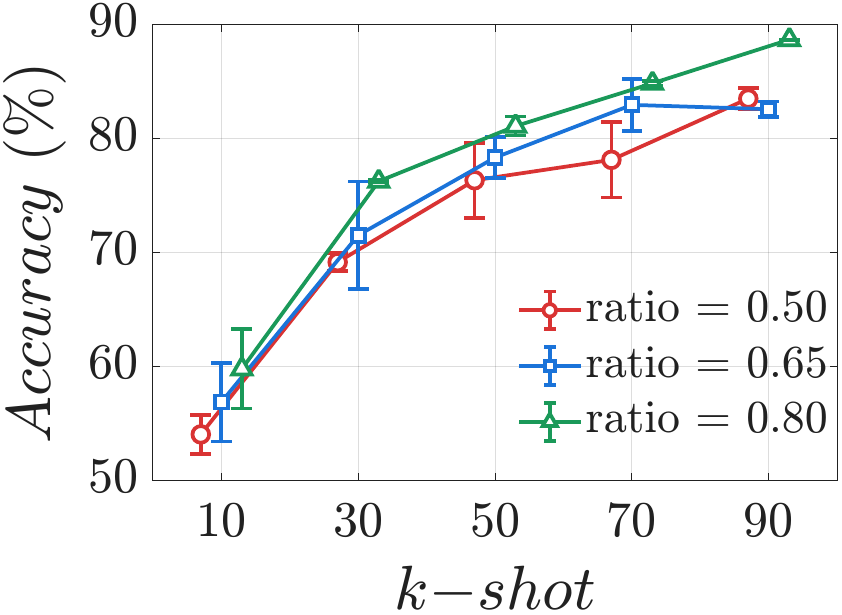}}\hspace{2mm}
\subfigure[NLOS (F1 Score)\label{fig:ratio_me_f1}]
{\includegraphics[width=0.48\columnwidth]{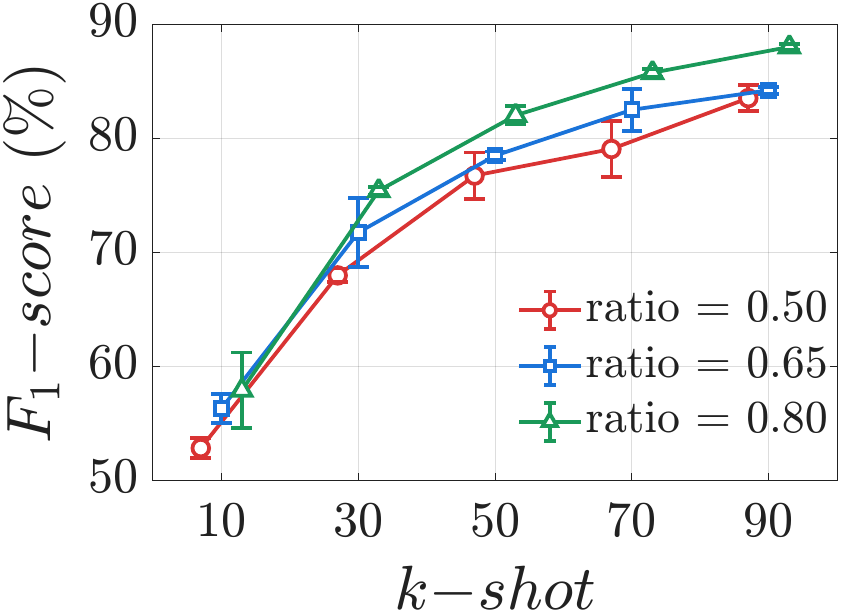}}
\vspace{1mm}
\subfigure[HTHI (Accuracy)\label{fig:ratio_hthi_acc}]
{\includegraphics[width=0.48\columnwidth]{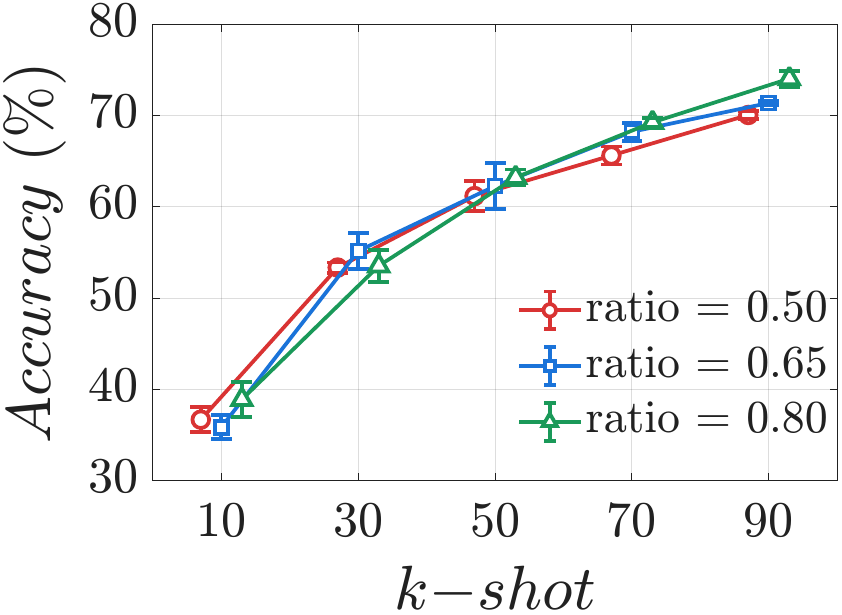}}\hspace{2mm}
\subfigure[HTHI (F1 Score)\label{fig:ratio_hthi_f1}]
{\includegraphics[width=0.48\columnwidth]{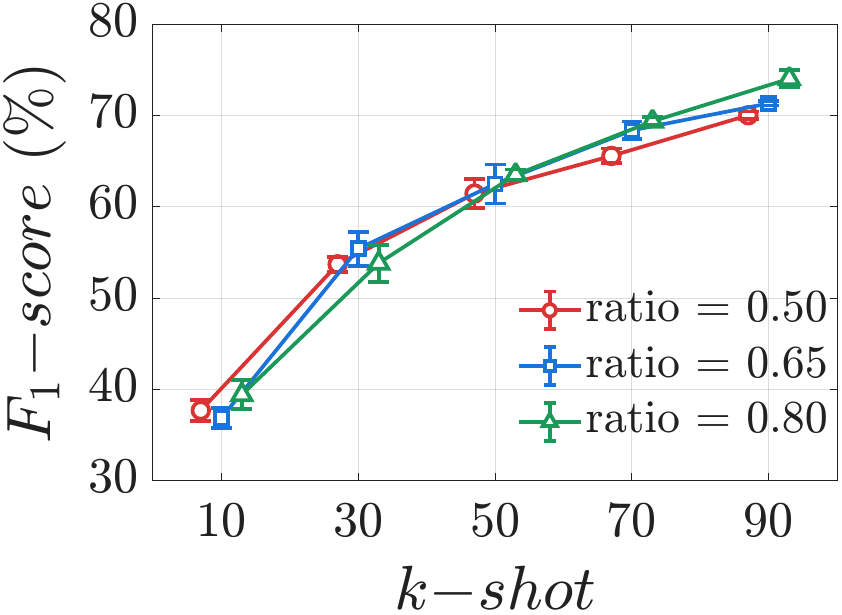}}
\caption{Impact of unlabeled pre-training data size and labeled fine-tuning data size on model performance. The x-axis represents the number of labeled samples per class used in $k$-shot learning, and different curves correspond to different pre-training data ratios (50\%, 65\%, 80\%).}
\label{fig:data_size_impact}
\end{figure}

We analyze the impact of labeled fine-tuning data size ($k$) and unlabeled pre-training data scale (50\%, 65\%, 80\%) in Fig.~\ref{fig:data_size_impact}.

First, performance monotonically improves with labeled samples ($k$) across all settings. This confirms that CIG-MAE learns a meaningful, linearly separable feature space where the classifier can effectively leverage additional supervision to refine decision boundaries.

Second, the impact of pre-training data volume depends heavily on dataset complexity. On the relatively clean SignFi dataset, increasing the unlabeled pre-training data from 50\% to 80\% yields only marginal gains, suggesting early saturation. This is likely because the SignFi subset used in this work (User 5 in a home environment) was collected in a comparatively controlled setting, so even 50\% of the unlabeled data already covers most of the dominant signal manifold.

By contrast, the benefit is much more pronounced on NLOS. In the NLOS 10-shot setting, for example, increasing the pre-training ratio from 50\% to 80\% improves the accuracy from 54.06\% to 59.80\%. This is consistent with the sensitivity analysis in Fig.~6, where NLOS requires a substantially larger BT projection dimension (5096 versus 1024 for SignFi/HTHI), suggesting a more complex representation structure. In addition, NLOS contains 10 subjects with distinct ages, weights, and heights, introducing greater subject-dependent variability. Together with stronger background interference and propagation uncertainty, this makes additional unlabeled data particularly valuable for learning invariant and noise-robust features.

HTHI shows an intermediate trend: the gain from more pre-training data is limited in the 10- and 30-shot settings, but becomes clearer from 50-shot onward. Since HTHI contains 40 interaction pairs randomly formed from 66 subjects with varying ages, weights, and heights, its representation space is also highly diverse. We conjecture that under extremely scarce supervision, the linear classifier is still the bottleneck, whereas with more labeled samples it can better exploit the improved pre-trained representation.
\subsection{Backbone Comparison and Analysis}
\label{backbone_comparison}
To validate the suitability of our design for practical IoT deployment, we benchmark the proposed CNN backbone against a ViT-based counterpart. Specifically, we replaced the CNN encoder-decoder with a ViT equivalent following the standard Masked Autoencoder design \cite{ViT,SSL_MAE_2021}, while keeping other components unchanged. The ViT baseline consists of an encoder with $6$ blocks ($1024$ embedding dim, $16$ heads) and a decoder with $5$ blocks ($512$ dim, $8$ heads), determined as the best-performing configuration by sequentially sweeping decoder depth over $\{1\text{--}6\}$, encoder depth over $\{1\text{--}7\}$, and embedding dimension over $\{256, 512, 1024\}$ on the NLOS dataset. The comprehensive comparison is detailed in Table~\ref{cnn_vit_compare}.

\begin{table}[htbp]
\centering
\caption{Performance and efficiency comparison between the proposed CNN backbone and the ViT baseline.}
\label{cnn_vit_compare}
\small
\resizebox{\columnwidth}{!}{
\begin{tabular}{l cc cc cc}
\toprule
\multirow{2}{*}{Method} & \multicolumn{2}{c}{SignFi} & \multicolumn{2}{c}{NLOS} & \multicolumn{2}{c}{HTHI} \\
\cmidrule(lr){2-3}\cmidrule(lr){4-5}\cmidrule(lr){6-7}
& Accuracy & F1 Score & Accuracy & F1 Score & Accuracy & F1 Score \\
\midrule
CNN (Ours) & \textbf{98.49 $\pm$ 0.27} & \textbf{98.42 $\pm$ 0.20} & \textbf{59.80 $\pm$ 3.48} & \textbf{57.93 $\pm$ 3.30} & \textbf{38.90 $\pm$ 1.93} & \textbf{39.43 $\pm$ 1.57} \\
ViT & $91.76 \pm 0.13$ & $91.09 \pm 0.01$ & $51.67 \pm 4.71$ & $43.00 \pm 2.26$ & $37.86 \pm 0.96$ & $36.14 \pm 0.01$ \\
\midrule
Params & \multicolumn{3}{l}{\textbf{CNN: 14.03M}} & \multicolumn{3}{l}{ViT: 152.08M} \\
FLOPs & \multicolumn{3}{l}{\textbf{CNN: 0.14G}} & \multicolumn{3}{l}{ViT: 129.98G} \\
Memory & \multicolumn{3}{l}{\textbf{CNN: 131.97MB}} & \multicolumn{3}{l}{ViT: 580.13MB} \\
\bottomrule
\end{tabular}
}
\end{table}

\textbf{Performance Superiority.} 
The CNN-based model substantially outperforms the ViT counterpart across all datasets. This gap stems from the inherent inductive biases of CNNs, such as locality and translation equivariance, which are highly compatible with the structured features of CSI time-frequency maps. In contrast, ViT lacks these priors and struggles to learn effectively from the limited scale of CSI datasets.

\textbf{Edge Deployment Feasibility.} Critically, the CNN architecture demonstrates superior efficiency, utilizing approximately $10\times$ fewer parameters and $1000\times$ fewer FLOPs than the ViT (Table~V). For practical context, commercial WiFi access points (e.g., Qualcomm IPQ4019 used by Nexmon
\cite{nexmon1,nexmon2,ipq}
) typically offer a general-purpose compute budget of only a few GFLOPs/s and 1\,GB of DRAM. Requiring about 0.14\,GFLOPs per inference, our CNN can theoretically sustain real-time sensing at roughly 20 FPS while comfortably fitting within the memory budget. In contrast, the ViT baseline's massive computational and memory demands render it entirely infeasible on such low-power SoCs without dedicated accelerators.

\textbf{Overfitting in Self-Supervised Pre-training.} Although the training splits contain only a few thousand unlabeled sequences (e.g., roughly $2.5\text{k}$ for SignFi), the masked autoencoding objective provides significantly more effective constraints than the raw sequence count suggests. Specifically, each sequence is partitioned into 400 patches. With a masking ratio of $\rho=0.95$, the model must reconstruct 380 patches per sequence. Over 300 pre-training epochs, this combinatorial effect yields approximately $2.5\text{k} \times 380 \times 300 \approx 2.8 \times 10^8$ patch-level predictions per stream. For our 14M-parameter CNN backbone, this equates to roughly ten supervisory constraints per parameter. Combined with convolutional weight sharing and the redundancy-reduction loss, this massive volume of fine-grained supervision acts as a strong regularizer. Consequently, increasing the pre-training data volume consistently improves test accuracy, confirming that overfitting is not dominant in the self-supervised stage.

\section{Discussion}\label{sec:discussion}
Although this work is evaluated on HAR, CIG-MAE is not inherently tied to activity semantics. Since CSI amplitude and phase jointly encode wireless propagation characteristics, the dual-stream masked reconstruction objective can also be viewed as learning propagation-aware priors from unlabeled channel observations. This creates a natural connection to broader wireless channel modeling tasks, such as radio map construction and channel twinning\cite{2026_RMConstruction_Tutorial,2023_SurveyDigitalTwins,2022_DigitalTwinofWirelessSystems}. In particular, AIM may help prioritize informative time-frequency regions for propagation modeling, while cross-modal alignment may support consistency across heterogeneous channel views\cite{williams2026digitaltwinassistedmeasurementdesign}. We emphasize that such extensions would require additional task-specific designs, such as spatial conditioning or map-level supervision, and are therefore beyond the scope of the present study\cite{luo2026wirelessdigitaltwincalibration}. From a practical perspective, resource-constrained deployment is also an important consideration. In this setting, the CNN backbone adopted in CIG-MAE provides a favorable basis for pruning, quantization, and distillation, while the pretrained encoder supports lightweight downstream adaptation\cite{2023_WiFiSensingontheEdge,wang2025cognitiveedgecomputingcomprehensive}. More broadly, a lightweight CSI-native pretraining framework for small-scale, noisy, and deployment-sensitive scenarios may also be relevant to future wireless foundation-model pipelines\cite{guler2025multitaskfoundationmodelwireless}.

\section{Conclusion}\label{sec:conclusion}
This paper proposes CIG-MAE, a generative self-supervised framework tailored for CSI-HAR. By integrating dual-stream masked reconstruction, AIM, and BT regularization, CIG-MAE effectively leverages the complementarity of amplitude and phase. This design eliminates the reliance on problematic data augmentations and negative samples while capturing intrinsic signal structures. Experiments on three public datasets demonstrate that CIG-MAE consistently achieves SOTA performance.  Critically, our comparison with ViT backbones confirms that the proposed CNN-based architecture offers superior accuracy with orders of magnitude lower computational cost ($1000\times$ fewer FLOPs), establishing a feasible path for deploying advanced sensing algorithms on resource-constrained edge devices. Future work will focus on several directions: evaluating scalability on larger, diverse datasets to further test generalization, and extending the framework to complex multi-person scenarios. Additionally, we plan to explore model compression techniques to further optimize real-time inference on commercial WiFi hardware and investigate optimization refinements for the discrete AIM policy, such as variance-reduction techniques, to further improve training efficiency.

\bibliographystyle{IEEEtran}
\bibliography{main}

\end{document}